%% file: arXivRev.tex
\documentclass[twocolumn,floatfix,amssymb,amsmath,secnumarabic,nofootinbib, nobalancelastpage,longbibliography]{revtex4-2}    

\usepackage[pdftex]{graphicx} 
\usepackage{dcolumn}  
\usepackage{bm}       
\usepackage[usenames,dvipsnames]{color}														
\definecolor{URLCOL}{rgb}{0,0.52,0.83} 
\definecolor{LINKCOL}{rgb}{0.05,0.5,0} 
\definecolor{CITECOL}{rgb}{0.25,0,0.48} 
\usepackage{epstopdf}
\usepackage[pdftex,bookmarks,breaklinks,bookmarksopen,bookmarksnumbered,colorlinks,linkcolor=LINKCOL,linktocpage,citecolor=CITECOL,urlcolor=URLCOL,pdfpagemode=UseOutline,pdftex]{hyperref}
\usepackage{datetime}

\def\preprintlink{ \href{http://dft.uci.edu + PAPER REF}{title of paper} }
\def\preprinttext{~}

\usepackage{fancyhdr}
\makeatletter
\fancypagestyle{titlepage}
{
	\lhead{\textsc{\preprinttext\ ~ \href{http://dft.uci.edu/publications.php}{~}}}
	\chead{}
	\rhead{\preprintlink}
	\lfoot{}
}
\makeatother
\def\preprintlink{ 
	\href{http://dft.uci.edu}
        {
~}
	}

\usepackage{graphicx}
\usepackage{amsmath}
\usepackage{physics} 
\usepackage{mathrsfs} 
\usepackage{subfigure} 
\usepackage{dcolumn} 
\usepackage{tabularx}
\usepackage{mathtools}
\usepackage{bm}
\usepackage{threeparttable}
\usepackage[normalem]{ulem}
\usepackage{array,multirow}
\usepackage{appendix}
\usepackage{epstopdf}
\usepackage{booktabs}
\usepackage{natbib}
\usepackage{feynmp}
\usepackage{threeparttable}
\usepackage{CJKutf8}

\definecolor{TITLECOL}{rgb}{0.1,0.2,0.7} 
\definecolor{PCOL}{rgb}{0.5,0.06,0.01} 
\definecolor{CHAPCOL}{rgb}{0,0.48,0} 
\definecolor{SECOL}{rgb}{0.1,0.2,0.7} 
\definecolor{CONTENTSCOL}{rgb}{0.1,0.2,0.7} 
\definecolor{SSECOL}{rgb}{0.25,0,0.48} 
\definecolor{SSSECOL}{rgb}{0.2,0.08,0.53} 
\definecolor{SHDCOL}{rgb}{0.4,0,0} 
\definecolor{ITMCOL}{rgb}{0.4,0,0} 
\definecolor{EXCOL}{rgb}{0,0.47,0.01} 
\definecolor{DEFCOL}{rgb}{0,0.42,0.01} 

\def\coloredtitle#1{\title{\textcolor{TITLECOL}{#1}}} 

\definecolor{URLCOL}{rgb}{0,0.17,0.43} 
\definecolor{LINKCOL}{rgb}{0.05,0.4,0} 
\definecolor{CITECOL}{rgb}{0.35,0,0.48} 

\definecolor{ngreen}{rgb}{0,0.48,0}

\def\sec#1{\section{\textcolor{SECOL}{#1}}}

\def\sectable#1{
\addcontentsline{toc}{subsection}{~~Table: \textcolor{SSECOL}{#1}}
\begin{table*}[h]
\caption{\bf \textcolor{SSECOL}{#1}}
}

\def\bea{\begin{eqnarray}}
\def\eea{\end{eqnarray}}
\def\ben{\begin{equation}}
\def\een{\end{equation}}
\def\benu{\begin{enumerate}}
\def\enu{\end{enumerate}}

\def\bei{\begin{itemize}}
\def\eei{\end{itemize}}
\def\beit{\begin{itemize}}
\def\eit{\end{itemize}}
\def\benu{\begin{enumerate}}
\def\enu{\end{enumerate}}

\def\n{n}

\def\sss{\scriptscriptstyle\rm}

\def\g{_\gamma}

\def\1var{(\bx_1...\bx)}

\def\half{\frac{1}{2}}

\def\bx{{x}}

\def\bj{{\bf j}}

\def\x{_{\sss X}}

\def\SD{^{\rm SD}} 

\def\sph_int{ {\int d^3 r}}

\definecolor{SPECOL}{rgb}{0,0.47,0.01}
\definecolor{QUOCOL}{rgb}{0,0,0.2}
\definecolor{SHDCOLb}{rgb}{0.69,0.4,0.1}

\definecolor{SPEQ}{rgb}{0.01,0.4,0.05} %

\definecolor{SPEQv}{rgb}{0.45,0.05,0.45} %

\definecolor{SPEQb}{rgb}{0.01,0.1,0.65} %

\definecolor{SPEQr}{rgb}{0.57,0.05,0.1} %

\def\sec#1{\section{\textcolor{SECOL}{#1}}}

\def\bay{\begin{array}}
\def\eay{\end{array}}
\def\bit{\begin{itemize}}
\def\beit{\begin{itemize}}
\def\eit{\end{itemize}}

\def\atan{\text{atan}}

\def\tanh{\text{tanh} }

\def\floor{\text{floor} }

\def\WKB{^\text{WKB}}
\def\CWKB{^\text{CWKB}}

\def\dd{~ \rotatebox{320}{\hspace{-5pt}\vbox to 5 pt {\hspace{-5pt} \hbox to 5pt {$\cdots$}}}\!\! }
\def\th{$\frac{1}{3}$}

\graphicspath{{./F/}}

\def\mbf{\mathbf}
\def\ra{\rightarrow}

\def\lf{\left}
\def\rg{\right}
\def\SWKB{^{\rm SWKB}}
\def\mc{\multicolumn}

\def\x{\times}
\def\CSWKB{^\text{CSWKB}}

\def\n{j} 
\begin{document}


\sf 
\coloredtitle{Asymptotics of eigenvalue sums when some turning points are complex}
\author{\color{CITECOL} Pavel Okun}
\affiliation{Department of Chemistry,
	University of California, Irvine, CA 92617,  USA}
\author{\color{CITECOL} Kieron Burke}
\affiliation{Departments of Physics and Astronomy and of Chemistry, University of California, Irvine, CA 92617}
\date{\today}
\begin{abstract}
Recent work has shown a deep connection between semilocal approximations in density functional theory and the asymptotics of the sum of the WKB semiclassical expansion for the eigenvalues.  However, all examples studied to date have potentials with only real classical turning points.  But systems with complex turning points generate subdominant terms beyond those in the WKB series.  The simplest case is a pure quartic oscillator.  We show how to generalize the asymptotics of eigenvalue sums to include subdominant contributions to the sums, if they are known for the eigenvalues.  These corrections to WKB greatly improve accuracy for eigenvalue sums, especially for many levels.  We obtain further improvements to the sums through hyperasymptotics.  For the lowest level, our summation method has error below $2 \times 10^{-4}$.  For the sum of the lowest 10 levels, our error is less than $10^{-22}$.   We report all results to many digits and include copious details of the asymptotic expansions and their derivation.
\end{abstract}


\maketitle
\def\floor#1{{\lfloor}#1{\rfloor}}
\def\sm#1{{\langle}#1{\rangle}}
\def\dis{_{disc}}
\newcommand{\Z}{\mathbb{Z}}
\newcommand{\R}{\mathbb{R}}
\def\w{^{(0)}}
\def\w{^{\rm WKB}}
\def\II{^{\rm II}}
\def\hd#1{\noindent{\bf\textcolor{red} {#1:}}}
\def\hb#1{\noindent{\bf\textcolor{blue} {#1:}}}
\def\eps{\epsilon}
\def\ew{\epsilon\w}
\def\ej{\epsilon_j}
\def\upet{^{(\eta)}}
\def\ejeta{\ej\upet}
\def\tjeta{\tj\upet}
\def\bej{{\bar \epsilon}_j}
\def\ewj{\epsilon\w_j}
\def\tj{t_j}
\def\vj{v_j}
\def\F{_{\sss F}}
\def\xt{x_{\sss T}}
\def\sc{^{\rm sc}}
\def\al{\alpha}
\def\ae{\al_e}
\def\bj{\bar j}
\def\bz{\bar\zeta}
\def\eq#1{Eq.\, (\ref{#1})}
\def\cN{{\cal N}}

\def\Lam{\Lambda}
\def\lam{\lambda}
\def\G{\Gamma}
\def\g{\gamma}
\def\eps{\epsilon}
\def\om{\omega}
\def\D{\Delta}
\def\d{\delta}
\def\r{\rho}
\def\a{\alpha}
\def\th{\theta}
\def\TH{\Theta}
\def\z{\zeta}
\def\b{\beta}
\def\sig{\sigma}
\def\Sig{\Sigma}

\sec{Introduction}
\label{sec:Intro}
This work grew out of an effort to derive systematic corrections to local density functional approximations.  The ultimate goal is to put density functional theory (DFT) on a similar theoretical foundation to wavefunction-based quantum chemistry, where it is well understood how to achieve greater accuracy (often starting from the Hartree-Fock approximation \cite{HH35}) with increasing computational cost \cite{SO96}.  The analagous starting point for density functional approximations are local density approximations but, because the expansion is an asymptotic semiclassical expansion, the mathematics required to perform this expansion is much less straightforward.

The connection between local density and semiclassical approximations has a long history \cite{MP56,E88}.  Work over the past 16 years has attempted to find that connection for the exchange-correlation (XC) energy of Kohn-Sham DFT \cite{PCSB06,EB09,BCGP16,CCKB18}, to provide insight for XC approximations \cite{SRP15,PRCV08}.  Recently, in simple one-dimensional systems, it has been possible to derive the leading corrections to local approximations from semiclassical expansions \cite{B20,B20b,BB20}. In typical cases, this expansion is asymptotic and so requires careful treatment \cite{B99,B91}.  The basic idea is that local density functional approximations are the leading-order terms in an asymptotic, semiclassical expansion and today's approximate functionals, such as generalized gradient approximations, are crude attempts to approximate higher order corrections.  Semiclassical quantum mechanics is widely discussed in the literature, especially in the context of dynamics \cite{H18}.  We refer readers interested in the application of semiclassics specifically to eigenstates to Ref. \cite{C14}.  We discuss the connection between semiclassics and DFT in Ref. \cite{OB21}.

We work directly with the total energy as a functional of the potential, but approximations can ultimately be converted into density functionals \cite{CLEB11,CGB13} for the 1D kinetic energy (whose three-dimensional counterpart is vital for orbital-free DFT \cite{LC05}).  The key is to approximate the sum of the first $N$ eigenvalues, $E_N$ (in chemistry parlance this is the ground state energy of $N$ non-interacting same-spin fermions).  The connection between this sum and density functionals is explained in Refs. \cite{B20b,OB21}.

The key idea is to generate an asymptotic expansion for the sum of eigenvalues directly from the known WKB series for individual eigenvalues.  Such asymptotics of sums are much more accurate than summing the asymptotic approximations for the individual eigenvalues.  Since this method for summing eigenvalues uses the Wentzel–Kramers–Brillouin (WKB) series \cite{BO99}, we call it SWKB, where the ''S" stands for sum.  It was developed in Refs. \cite{BB20,B20,B20b}.  In simple cases (particle in a box and harmonic oscillator), the semiclassical expansion truncates at low order, yielding exact results for both the eigenvalues and their sums \cite{B20b}.  Reference \cite{B20} analyzed the P{\"o}schl-Teller well, where the WKB and SWKB series are convergent.  Reference \cite{BB20} tested the SWKB approximation on a linear potential with a hard wall (the linear half-well), which is truly asymptotic.

The P\"{o}schl-Teller and linear wells have only real classical turning points, i.e., real values of $x$ where $v(x) = \eps$.  But many other simple analytic forms also allow for complex turning points \cite{BJ12}, that is, complex solutions to $v(x) = \eps$ with finite imaginary values.  Such complex turning points produce exponentially small corrections that are missed by the standard WKB expansion.  The quartic oscillator is the simplest, and it has long been known to have sub-dominant contributions to its semiclassical eigenvalue expansion, which reduce errors significantly relative to WKB to the same order.  So here we apply the asymptotic summation method to the quartic oscillator.  We show how the subdominant corrections to the WKB eigenvalues lead to subdominant corrections to the SWKB sums, and generate an asymptotic expansion for these corrections.  We find that the subdominant corrections to the sums likewise greatly improve accuracy.  This methodology can be easily applied to any such case.

More broadly, the quartic oscillator is of interest in particle physics for testing model field theories \cite{DGKS93,CKS95}.  In mathematical physics it is a model asymptotic system \cite{ABS19}.  The quartic oscillator Schr{\"o}dinger equation is a special case of Huen's differential equation (Sec. 31.2 of Ref. \cite{DLMF}) and its exact solution was studied in this context in Refs. \cite{L97,BL97}.  While we focus on the asymptotic approximation to all eigenvalues Ref. \cite{LMT06} uses a different technique to approximate the ground state wavefunctions and energies.  The anharmonic oscillator, with both quartic and harmonic terms in the potential \cite{LMSW69,S69,BW69,GGS70}, has generated more interest in the physics community than the pure quartic oscillator studied here.  The pure quartic oscillator represents the strong coupling limit of the anharmonic oscillator \cite{HM75}.  We use it here as the simplest case to test summation techniques in the presence of complex turning points.

In Sec. \ref{sec:EigVal} we thoroughly analyze the semiclassical expansion of the quartic oscillator eigenvalues to many orders.  While the basic structure of this expansion (WKB plus subdominant corrections) has long been known \cite{BPV79,V83,BPV78}, we examine it more thoroughly and systematically than typical studies in the literature.  We find asymptotic expansions beyond 20 orders, subdominant corrections to such expansions, 20 eigenvalues given to 40 digits, and in some cases errors less than $10^{-20}$.  We also give explicit formulas for the exact analytic expressions for the lower-order asymptotic coefficients, and the formulas to generate the exact WKB series and its inversion to all orders, making it easy for others to use our work.

But all this is a prelude to the main purpose of our work, which is to apply newly developed methods of finding {\em sums} of the lowest $N$ eigenvalues of quantum problems.  In Sec. \ref{sec:Sum} we asymptotically sum over the explicit WKB expansion of Sec. \ref{sec:EigVal}, finding its asymptotic coefficients.  The crucial point is that, from the subdominant corrections to WKB, we can easily find the subdominant corrections to the WKB asymptotic expansion for the sums, which are far more accurate than their WKB counterparts.  We also use tricks of hyperasymptotics for the sums, to further improve accuracy.  This is the most accurate approximation which we derive.  Finally, we take differences between approximations for sums to approximate the individual eigenvalues, which we can again compare to the WKB series including subdominant corrections.

While our results should be of general interest to anyone studying semiclassical expansions of eigenenergies in any context, our driving interest is in density functional theory (DFT).  Thus in Sec. \ref{sec:Con} we summarize our findings and explain their relevance to DFT.

\sec{Notation and conventions}

We choose units so that our Hamiltonian is
\begin{equation}
\hat{\mathscr{H}} = - \half \frac{d^2}{d x^2} + \frac{x^4}{2}.
\end{equation}
Many benchmark results to many digits were reported in Ref. \cite{OB21b}, with a different constant.  So we give the first 20 eigenvalues to 41 decimal places in Table S1.  References \cite{B19,BO99,R70,BOW77} also report eigenvalues for this system, but to less digits.

We adopt the following notation for asymptotic series.  Define the function
\begin{equation}
B_M(y) = \sum_{m=0}^M b_m(y) = \sum_{m=0}^M \frac{b_m}{y^m},
\end{equation}
i.e., given a set of coefficients $b_m$, $B_M(y)$ is the $M$-th order polynomial in $1/y$.  We choose $b_0=1$ always, so that $b_0(y)=1$ and $B_M(y) \to 1$ as $y\to\infty$.  Thus the dominant asymptotic behavior is always given by setting the polynomial to $1$.  We refer to each $b_m(y)$ as the $m$-th addition, rather than the $m$-th term, as the smallest magnitude term is traditionally called the least addition in the field of asymptotics.

We will often use optimal truncation to create approximations that are more accurate than any of fixed order (superasymptotics) \cite{B91}.  We consider three different choices.  Empirical ($E$) optimal truncation implies truncation at the order yielding the lowest error.  This requires knowing which function is being expanded.   Least-addition ($L$) optimal truncation means truncating at that value of $m$ for which $|b_m(y)|$ is least.  Finally, asymptotic ($A$) optimal truncation uses the (usually simple but not always available) formula for least addition as $m\to\infty$, using the asymptotically dominant expression for $b_m(y)$ as $m\to\infty$. All three must agree for sufficiently large $y$, but may disagree for moderate values of $y$.  We generate at least 7 different asymptotic expansions in this paper, and for ease of use the coefficients of the most important expansions are given in Table \ref{tab:EigCoeff} of Appendix \ref{sec:TabConst}  to 13 digits, but 41 digits are given in the supplemental info.

Throughout, we define error as the difference between approximate and exact.

\sec{Semiclassical expansion of quartic oscillator eigenvalues}
\label{sec:EigVal}

In this section, we thoroughly review the known WKB series for the quartic oscillator, and its leading subdominant corrections.  All the essential facts already appear in various places in the literature \cite{BO99,BPV79,BPV79,V83,BPV78,BOW77}, but we collect the relevant ones here and are more thorough.

The WKB series for the quartic oscillator eigenvalues is well-known:
\begin{equation}
\label{WKBImp}
\lf(\frac{\eps}{\a}\rg)^{3/4} A_M[(2\eps)^{3/2}] = z,
\end{equation}
where $\g = \G(1/4) \approx 3.62561$, $\a = 3\pi^{2}[3/(2\g^8)]^{1/3} \approx 1.09253$ and eigenvalues $\eps_\n$ occur at $z=\n+1/2$, $\n=0,1,2,\cdots$.  We derive Eq. (\ref{WKBImp}) in Appendix \ref{sec:DeriveImpWKB}.  The first 6 $a_n$ are reported in Bender and Orszag (where $A_{2n} = A_0 a_n$).  We give the $a_n$ exactly up to $n = 28$ in Table S3.  Like many of the series in this paper, Eq. (\ref{WKBImp}) is an asymptotic expansion and never converges.  The coefficients at first become smaller, but after $m = 2$, they grow rapidly, with asymptotic behavior \cite{BPV79}
\begin{equation}
\label{AAsymp}
a_n^{(0)} =  - (-1)^{\floor{n/2}} \frac{2}{\pi} \left(\frac{9\pi}{\g^4}\right)^n(2n-2)!, \qquad n \ra \infty.
\end{equation}
Their signs also oscillate, with pairs of positive values followed by pairs of negative values.  The $a_n$ appear in Table \ref{tab:EigCoeff} and to more digits in Table S2.  This table also compares the asymptotic approximation in Eq. (\ref{AAsymp}) with the exact $a_n$.

\input{TableEigValApprox}

Equation (\ref{WKBImp}) yields an implicit definition of $\eps(z)$ at each order $M$.  We can invert the expansion order-by-order to find
\begin{equation}
\label{WKBExpl}
\eps\WKB_M(z) =  \a\, z^{4/3}\, B_M(z^2),
\end{equation}
where the coefficients $b_n$ are also listed numerically in Table \ref{tab:EigCoeff} and given to more digits in Table S5.  Since Eq. (\ref{WKBExpl}) is the inversion of Eq. (\ref{WKBImp}) the $\{b_n\}$ are linear combinations of the $\{a_{n'}\}$ with $n' \leq n$.  We discuss how to generate the $b_n$ and their exact forms  in Appendix \ref{sec:AnalyticStructureCoeff}.  Assuming that the dominant contribution to the $b_n$ is the term with the highest order $a_n$ in Eq. (\ref{Rec.b}) yields
\begin{equation}
\label{BAsymp}
b_n^{(0)} = (-1)^{\floor{n/2}} \frac{8(2n-2)!}{3\pi(2\pi^2)^n}, \qquad n \ra \infty.
\end{equation}
Table S5 gives the ratio of $b_n$ coefficients to this asymptotic form, and strongly suggests it is correct.

\begin{figure}[htb!]
\includegraphics[width=0.8\columnwidth]{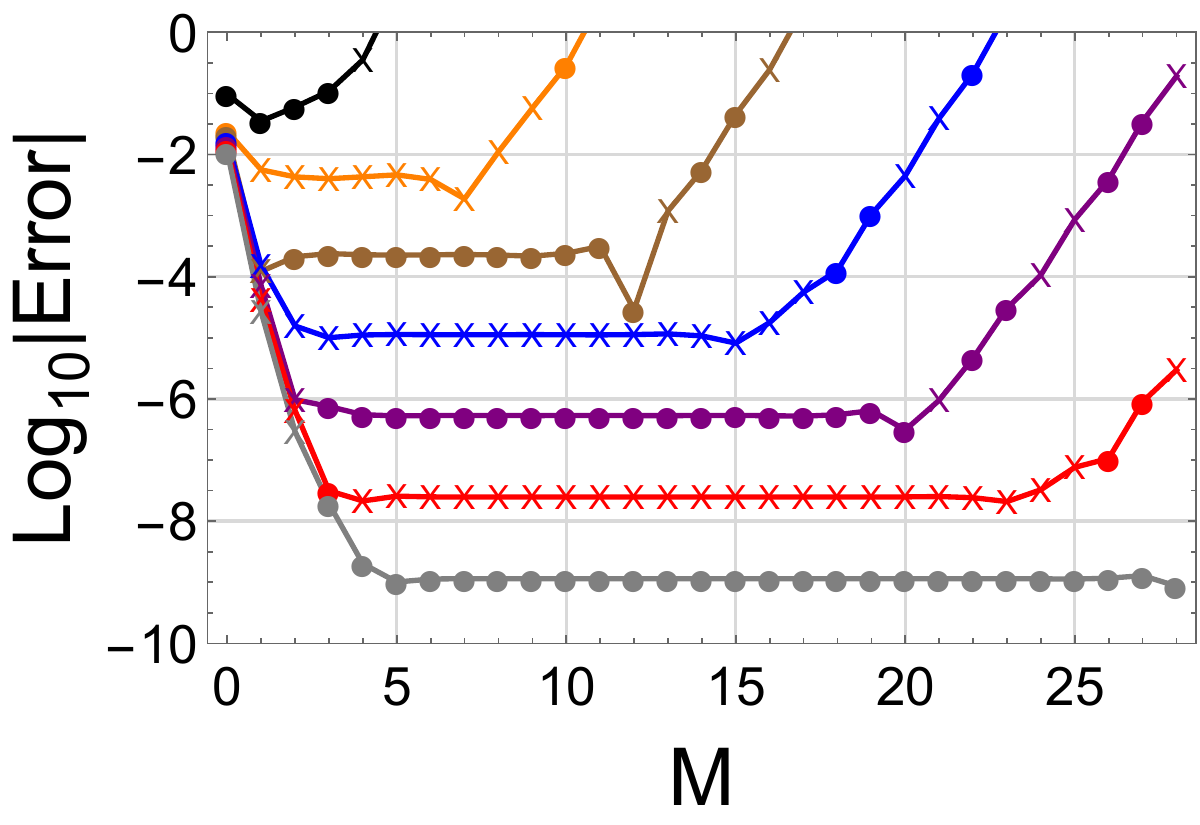}
\caption{Errors (approximate minus exact) of the WKB eigenvalue series as a function of order, $M$, for the ground state (black) to the sixth excited state (gray).  The sign of the error is denoted by the symbol (X's positive, full circles negative).}
\label{fig:WKBEig}
\end{figure}

Figure \ref{fig:WKBEig} plots the errors for Eq. (\ref{WKBExpl}) as a function of order for the first 7 eigenvalues.  While the WKB series is extremely accurate and improves as $n$ increases, plateaus develop of (approximately) fixed error, and these widen as $n$ increases.  Thus there is a range of values of $M$ that all produce comparable error.  On the other hand the WKB series for the linear half well in Refs. \cite{B20b,BB20} smoothly decreases with order down to some minium value before smoothly increasing with increasing order.

\begin{figure}[htb!]
\includegraphics[width=0.8\columnwidth]{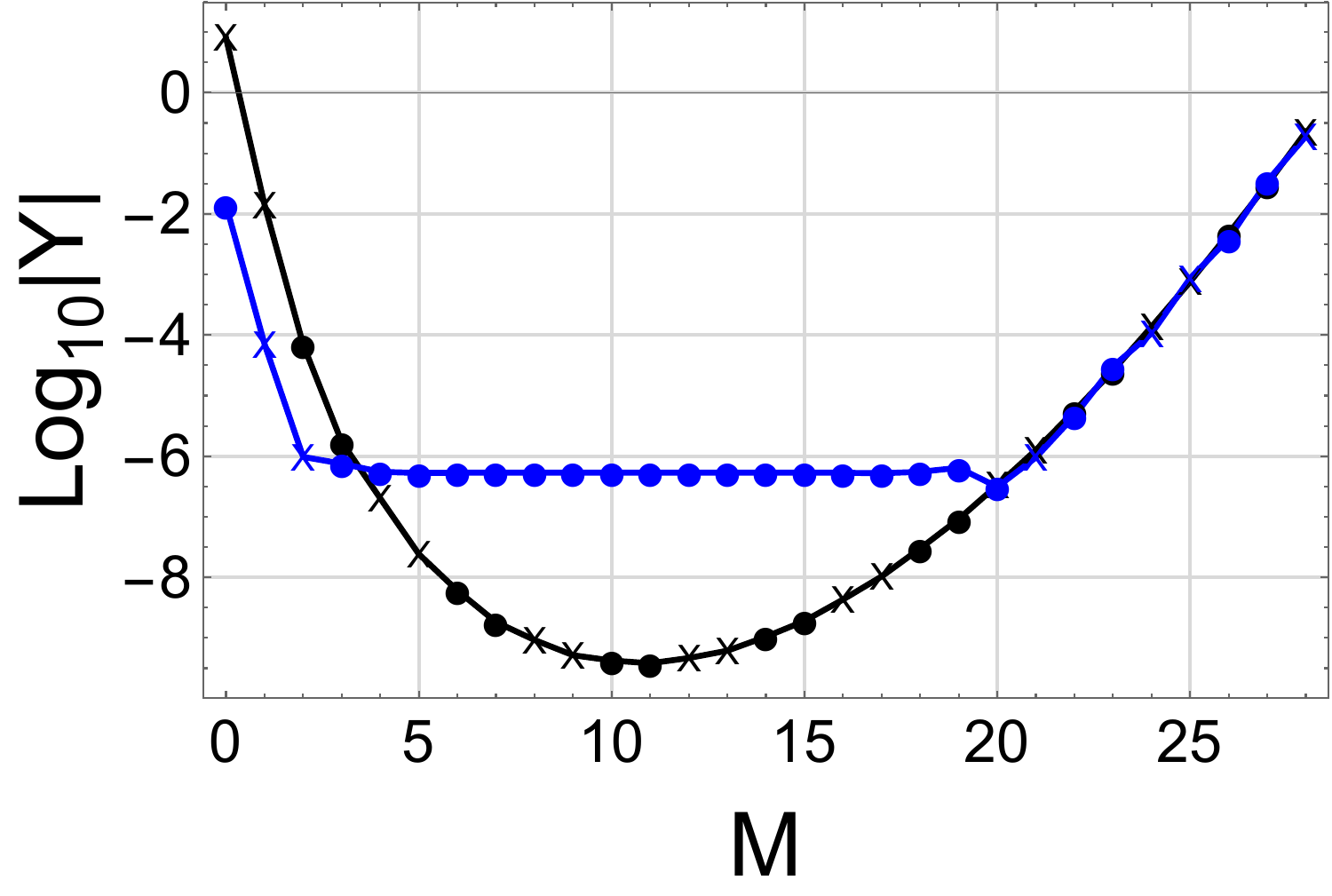}
\caption{WKB expansion errors (blue) and additions (black) for $\n = 4$, with conventions of Fig. \ref{fig:WKBEig}.}  
\label{fig:Eig4WKB}
\end{figure}

To see this in more detail we consider the fourth excited state ($\n = 4$) in Fig. \ref{fig:Eig4WKB} and include the magnitude of the additions [$\eps\WKB_M(z) - \eps\WKB_{M-1}(z)$].  On the plateau, the additions become much smaller than the error.  To the right, the additions grow and dominate the error.  This strongly suggests that something is missing from the WKB series itself.

For this problem, the WKB series is known not to give the complete semiclassical expansion of the eigenvalues.  The WKB series accounts for only the real turning points, but there are also two imaginary turning points in the complex plane.  For the linear half well \cite{BB20} the WKB series gave the full semiclassical expansion of the eigenvalues because there was only one real turning point (the hard wall only kills off the even states of the linear well and does not count as a regular turning point \cite{OB21}).  For the quartic oscillator, the true semiclassical expansion has additional subdominant pieces which the WKB series neglects.  Voros et. al \cite{BPV79} found a quantization rule that, unlike Eq. (\ref{WKBImp}), which considers only real turning points, takes all four turning points (both real and imaginary) into account:
\begin{equation}
\label{SCQuant}
z = (\eps/\a)^{3/4} A_M [(2\eps)^{3/2}] - g_{\n M} (\eps),
\end{equation}
where
\begin{equation}
\label{g}
g_{\n M}(\eps) = \frac{(-1)^\n}{\pi} \atan \exp \lf[ - \pi (\eps/\a)^{3/4} A_M [-(2\eps)^{3/2}] \rg].
\end{equation}
This more general quantization rule \cite{BPV79,V83,BPV78,V80,DP97} comes from the complex WKB method \cite{WK17,BW69} which extends the WKB expansion into the complex plane.

Noting that $g$ is much smaller than any power of $z$ in the large $z$ limit, the net effect of $g$ is simply to shift the $z$ value.  Then repeating the inversion which gave us Eq. (\ref{WKBExpl}) from Eq. (\ref{WKBImp}), we find
\begin{equation}
\label{SCFromWKB}
\eps_{M M'}(z) = \eps\WKB_M[z+g_{M'}(z)].
\end{equation}
Recalling that $\n = z - 1/2$, we rewrite Eq. (\ref{g}) as
\begin{equation}
\label{gExp}
g_{M'}(z) = \frac{(-1)^\n}{\pi} \atan~ \exp[-\pi\,z\, D_{M'}(z^2)],
\end{equation}
where we find the $d_n$ by plugging the WKB series in Eq. (\ref{WKBExpl}) into Eq. (\ref{g}) and rearranging the double sum into a single sum by grouping terms of like powers in $z$.  The $d_n$ are reported in Table \ref{tab:EigCoeff} and more precisely in Table S8.  Analytic expressions for the $d_n$ are given in  Appendix \ref{sec:AnalyticStructureCoeff}.

\begin{figure}[htb!]
\includegraphics[width=0.8\columnwidth]{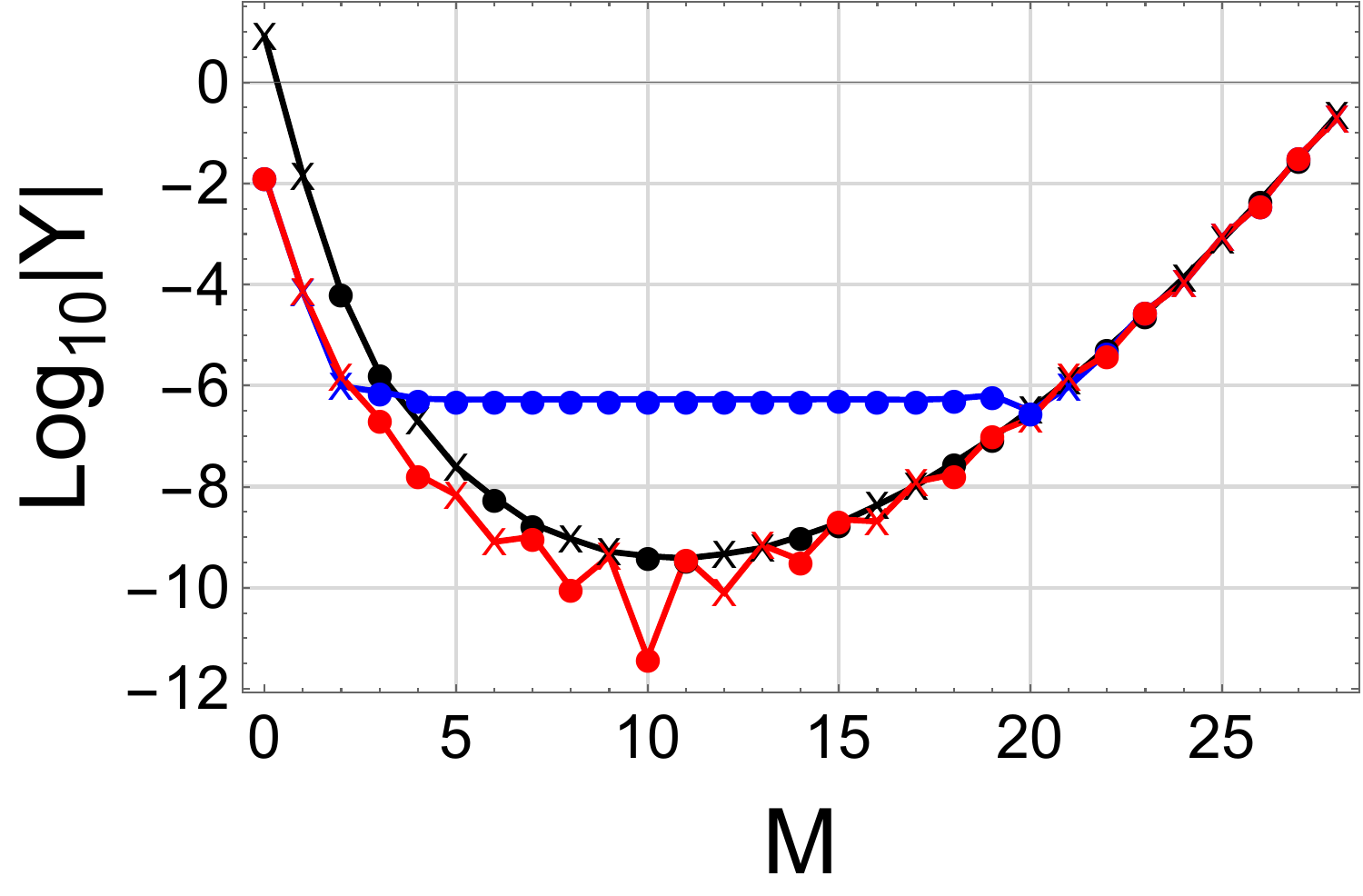}
\caption{Same as Fig. \ref{fig:Eig4WKB}, but adding the BCWKB (red) approximation of Eq. (\ref{CWKB}), using subdominant corrections truncated via least addition}.
\label{fig:WKBvSC}
\end{figure}

There is no necessary connection between $M$ and $M'$ in Eq. (\ref{SCFromWKB}): each can be chosen independently.  Here $M$ is the order at which we truncate the WKB series, and has a large effect on the accuracy of our approximations beyond WKB.  Thus we choose $M$ very carefully.  The order $M'$ determines the subdominant (SD) correction to the WKB values.  This is the first of several series we will derive for the SD values.  We use least addition to choose the order for all these SD series.  Varying $M'$ around the order of least addition has little effect on accuracy.  We label this BCWKB, for bare corrected WKB:
\def\BCWKB{^{\rm BCWKB}}
\begin{equation}
\label{CWKB}
\eps\BCWKB(z) = \eps\WKB_M[z + g_L(z)],
\end{equation}
where $g_L(z)$ is optimally truncated via least addition.  Equation \ref{CWKB} yields our best estimate for the individual eigenvalues, Table S18 shows the $D_L(z^2)$ and the values of $L$, which we find to be $L=2 \n + 4$ for all $\n > 2$.  We plug these $D_L$ into Eq. (\ref{gExp}) to estimate $g$ for each $\n$, and then evaluate $\eps\BCWKB(z)$.  This yields the corrections shown in Fig. \ref{fig:WKBvSC}. 

\begin{figure}[htb!]
\includegraphics[width=0.8\columnwidth]{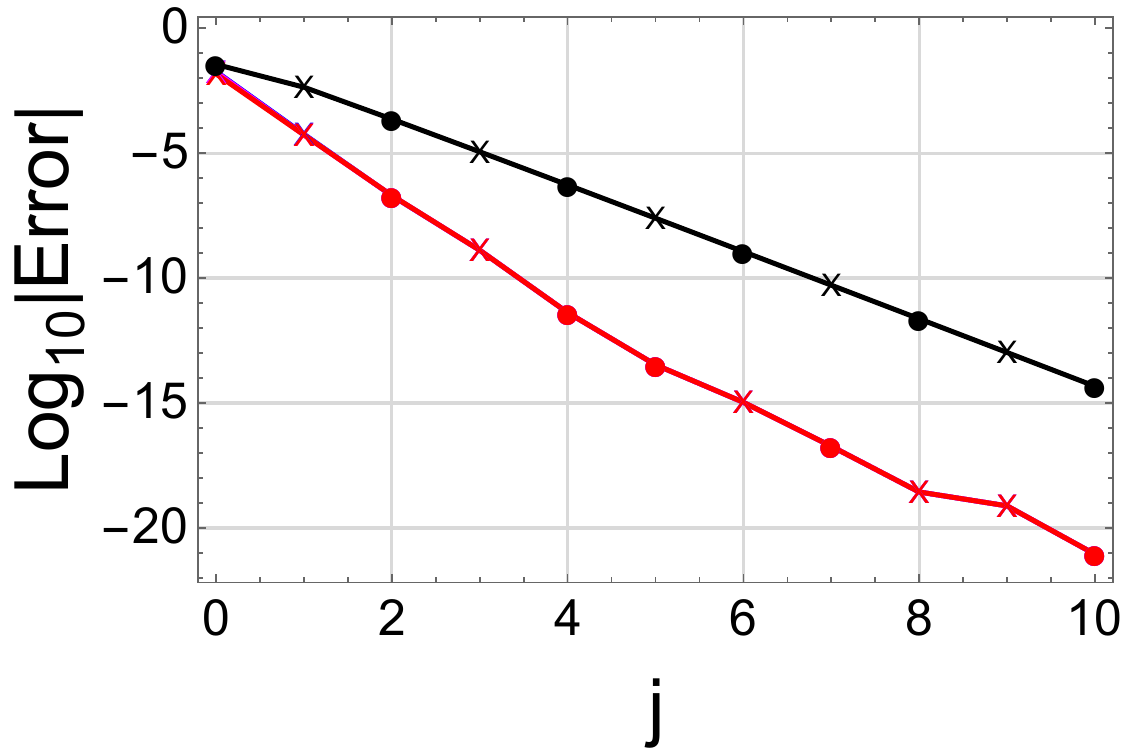}
\caption{WKB (black) and corrected WKB errors (red) after optimal truncation, from Table \ref{tab:SCError}.  All different corrected approximations (BCWKB, $\eps_M\WKB + \D\eps_L$, CWKB) are indistinguishable.  Conventions as in Fig. \ref{fig:WKBEig}.}
\label{fig:EigvalTableErr}
\end{figure}

This figure shows extraordinary improvement over the bare WKB series.  To determine the asymptotic ($\n \ra \infty)$ order of least addition for the WKB series, we approximate the $b_n$ with Eq. (\ref{BAsymp}) and minimize the terms, the $b_n(z^2)$, in Eq. (\ref{WKBExpl}) with respect to $M$:
\begin{equation}
\label{AsymTrunc}
A(\n) = \bigg\lfloor 1 + \frac{\pi}{\sqrt{2}}\left( \n + \half \right) \bigg\rfloor.
\end{equation}
The asymptotic and least addition orders of optimal truncation are identical for $\n \geq 5$.  However, the empirical least error is one order less, for reasons we have been unable to fathom.  The figure, made for $j=4$, has $L=11$, but the actual least error is at $M=10$, by almost two orders of magnitude.  Table \ref{tab:SCError} documents the enormous improvement of Eq. (\ref{CWKB}) over WKB.  We plot the errors in this table in Fig. \ref{fig:EigvalTableErr}.  For the first level, $L=2$, and the improvement is only a factor of 2.  For $\n=10$, the improvement is 7 orders of magnitude.  The prefactor of $\n$ in Eq. (\ref{AsymTrunc}), $\pi/\sqrt{2}$, explains why our accuracies are lower than those of Ref. \cite{BB20}.  For both eigenvalues and their sums in the next section, logs of errors are $1/\sqrt{2}$ smaller than those of Ref. \cite{BB20} as $\n \ra \infty$.

Last in this section, we find the asymptotic behavior of the subdominant corrections to WKB.  We expand around $g=0$, i.e., the WKB series, to find the difference with WKB to be
\begin{equation}
\label{SC1}
\D \eps_M(z) \approx g_L(z) \frac{d\eps\WKB_M}{dz} = g_L(z) \a z^{1/3} B_M'(z^2),
\end{equation}
where $b_m' = (4/3-2m)b_m$.  In Table \ref{tab:SCError}, we show that adding this to the WKB series yields results about as accurate as those of Eq. (\ref{CWKB}).  Now our expression is linear in $g(z)$.   We next expand all functions in $g(z)$ in inverse powers of $1/z$, {\em except} for the dominant exponential, $\exp(-\pi z)$, yielding
\begin{equation}
\label{eSD}
\eps\SD_M(z) = (-1)^\n \frac{4\a}{3\pi}  z^{1/3} e^{-\pi z} H_M(z),
\end{equation}
where the superscript SD refers to the leading subdominant contribution.  We define the corrected-WKB (CWKB) approximation:
\begin{equation}
\label{DefCWKB}
\eps\CWKB_M(\n) = \eps\WKB_M(\n + 1/2) + \eps\SD_L(\n + 1/2),
\end{equation}   
where the SD corrections are least-addition optimally truncated.  For $\n > 2$, we find $L =4\n + 7$ (Tables \ref{tab:EigvalSD} and S19).   In Table \ref{tab:SCError} we show that Eqs. (\ref{CWKB}), (\ref{SC1}), and (\ref{eSD}) all produce results of comparable accuracy, with differences becoming negligible with increasing level.  We give these subdominant energies in Table \ref{tab:EigvalSD} and more fully in Table S19.  We give the $h_n$ numerically in Table \ref{tab:EigCoeff}, more precisely in Table S11, and analytically in Appendix \ref{sec:AnalyticStructureCoeff}.  Equation (\ref{DefCWKB}) yields about the same errors as Eq. (\ref{SC1}) in Table \ref{tab:SCError} to the number of digits shown in that table.  In Table S21 we show that WKB is relatively insensitive to the order of truncation, but CWKB is extremely sensitive, so we choose $M = L - 1$ in Eq. (\ref{DefCWKB}).

\begin{table}[!ht]
$\begin{array}{|c|c|r|}
\hline
\n & L & \mc{1}{c|}{\eps_L\SD(\n + 1/2)} \\
\hline
0 & 2  &  4.98423055821\x 10^{-02} \\
1 & 7  & -4.25176939432\x 10^{-03} \\
2 & 11 &  2.28025923533\x 10^{-04} \\
3 & 19 & -1.12467805687\x 10^{-05} \\
4 & 23 &  5.34360714255\x 10^{-07} \\
5 & 27 & -2.48624007875\x 10^{-08} \\
6 & 31 &  1.14140669578\x 10^{-09} \\
7 & 35 & -5.19166517060\x 10^{-11} \\
8 & 39 &  2.34538047637\x 10^{-12} \\
9 & 43 & -1.05403481095\x 10^{-13} \\
10 & 47 &  4.71745228714\x 10^{-15} \\
\hline
\end{array}$
\caption{Subdominant corrections to WKB eigenvalues of Eq. (\ref{eSD}), truncated at least addition, L.}
\label{tab:EigvalSD}
\end{table}

Our treatment of the subdominant corrections to the WKB series is simpler than that employed in modern particle theory \cite{DU14,DU17}, but suffices for our study of the asymptotic approximation to the eigenvalue sum.  For those interested in the techniques of modern asymptotic theory and its application to quantum oscillators we refer them to Ref. \cite{DDP97}.

\sec{Asymptotics of sums}
\label{sec:Sum}

We now turn to the main topic of this work, the asymptotics of sums of eigenvalues, and whether or not subdominant contributions can also be incorporated in such approximate sums.  Given the extraordinary accuracy of modern DFT approximations (energy differences can be accurate to within one part in 10$^9$ of total energies in Kohn-Sham (KS) calculations), the asymptotic DFT for model systems developed in Refs. \cite{B20,B20b,BB20} might not be sufficient in the presence of subdominant corrections.  It is also of intellectual interest to see if they can be handled.

\begin{figure}[htb!]
\includegraphics[width=0.8\columnwidth]{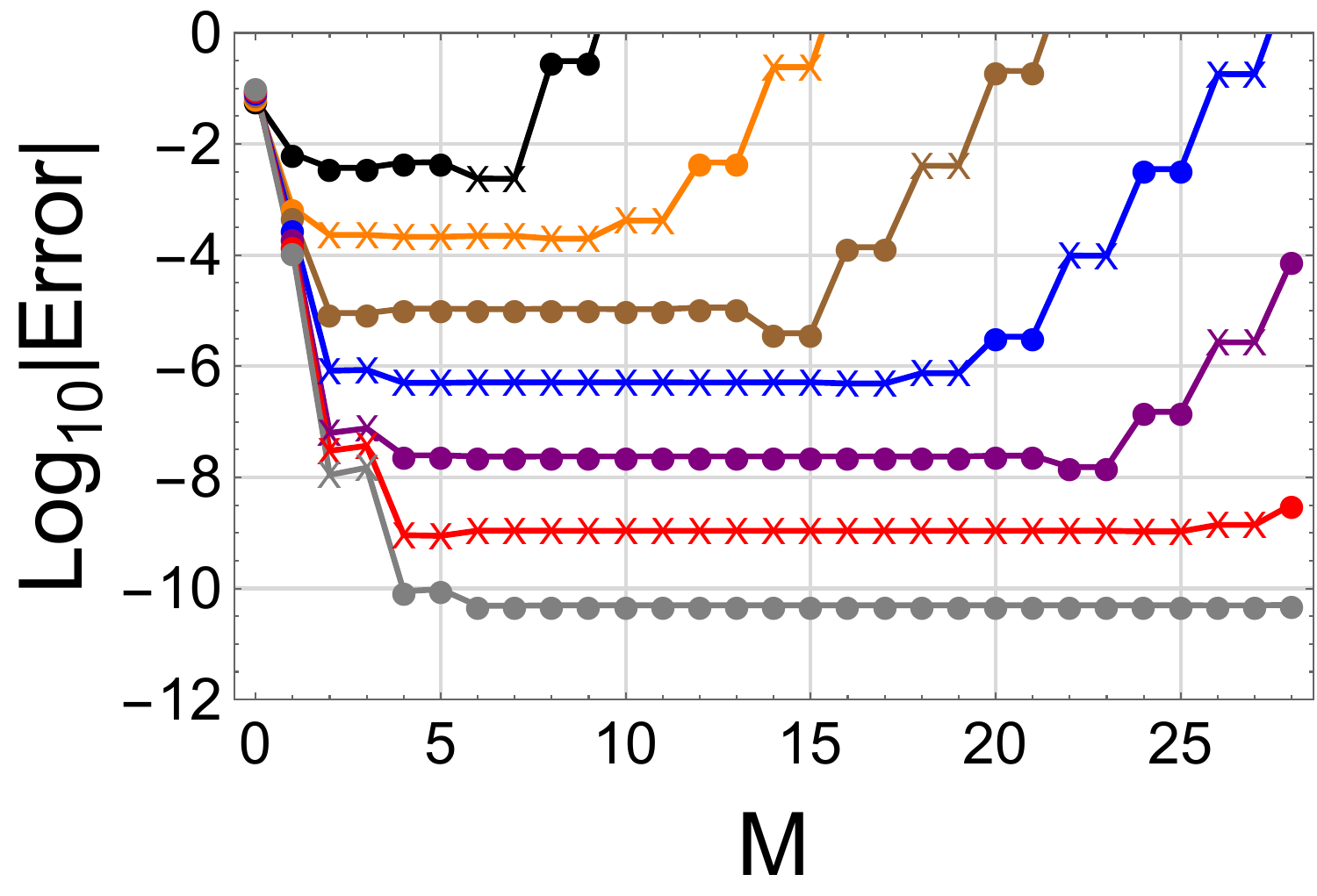}
\caption{Errors in WKB approximation for sums (SWKB) from $N = 1$ (black) to $N = 7$ (gray), conventions as in Fig. \ref{fig:WKBEig}.}
\label{fig:BBErr}
\end{figure}

The sum of the first $N$ eigenvalues is
\begin{equation}
\label{ExEng}
E(N) = \sum_{\n = 0}^{N-1} \eps_\n.
\end{equation}
We can derive the SWKB approximation to this sum using any of the methods in Refs. \cite{B20b,B20,BB20}.  We choose to start with Eq. (2.4) of Ref. \cite{BB20}:
\begin{equation}
\label{SumForm}
E_R(N) = E(\infty) - \int_{N}^{\infty} dj\, \eps_j - \frac{\eps_N}{2} + \sum_{r = 1}^{R} \frac{B_{2r}}{(2r)!} \lf(\frac{d}{dN}\rg)^{2r-1}\eps_N,
\end{equation}
where the $B_{2m}$ are the Bernoulli numbers and $E(\infty)$ is the carefully regularized sum over all eigenvalues (here $E(\infty) = 0$).  We insert the WKB expansion for $\eps_j$, Eq. (\ref{WKBExpl}), into Eq. (\ref{SumForm}).  Equation (2.6) in Ref. \cite{BB20} explains how to evaluate the integral in Eq. (\ref{SumForm}):
\begin{equation}
\label{SumFormIntegral}
- \int_{N}^{\infty} dz\, \eps\WKB_M(z) = \a \sum_{m = 0}^{M} \frac{b_m}{7/3-2m} Z^{7/3-2m},
\end{equation}
where $Z = N + 1/2$.  Noting that 
\begin{equation}
\lf(\frac{d}{dN}\rg)^q Z^{p+q} = Z^p (p + 1)_q,
\end{equation}
where $(x)_n = \G(x + n)/\G(x)$ is the Pochhammer symbol (Sec. 5.2 of Ref. \cite{DLMF}), we find
\begin{equation}
\label{GExpan}
E_M(N) = \a \sum_{m = 0}^{M} \lf( \frac{3}{7} q_m Z - \frac{b_m}{2} \rg) Z^{4/3-2m},
\end{equation}
where
\begin{equation}
\label{q_of_b}
q_n = \frac{7/3}{7/3-2n} \sum_{m = 0}^{n} B_{2(n-m)} \binom{4/3-2m}{2(n-m)} b_m,
\end{equation}
where we use the binomial coefficient: $\binom{x}{r} = x...(x-r+1)/r!$.  The structure of Eq. (\ref{GExpan}) is analyzed in Appendix \ref{sec:NHalf}.  Expanding in powers of $N^{-1}$ yields
\begin{equation}
\label{PQO.BB}
E_M(N) = \frac{3}{7} \a  N^{7/3} C_M(N^2),
\end{equation}
where
\begin{equation}
\label{Form.C}
c_n = \frac{7}{7 - 6n} \sum_{m=0}^{n} \binom{2n - 7/3}{2m} K_m b_{n-m},
\end{equation}
with $K_0 = 1$ and
\begin{equation}
\label{Const.C}
K_m = \frac{1 - 4^{1-m} m}{2} + \sum_{r = 1}^{m} \left( \frac{1}{4^r} - \frac{1}{1+2r} \right)  \binom{2m}{2r} B_{2(m-r)},
\end{equation}
for $m > 0$.  We give the $c_n$ to 41 decimal places in Table S14.  The errors of Eq. (\ref{PQO.BB}) appear in Fig. \ref{fig:BBErr}, showing missing subdominant corrections.

\input{TableSumApprox}

\begin{figure}[htb!]
\includegraphics[width=0.8\columnwidth]{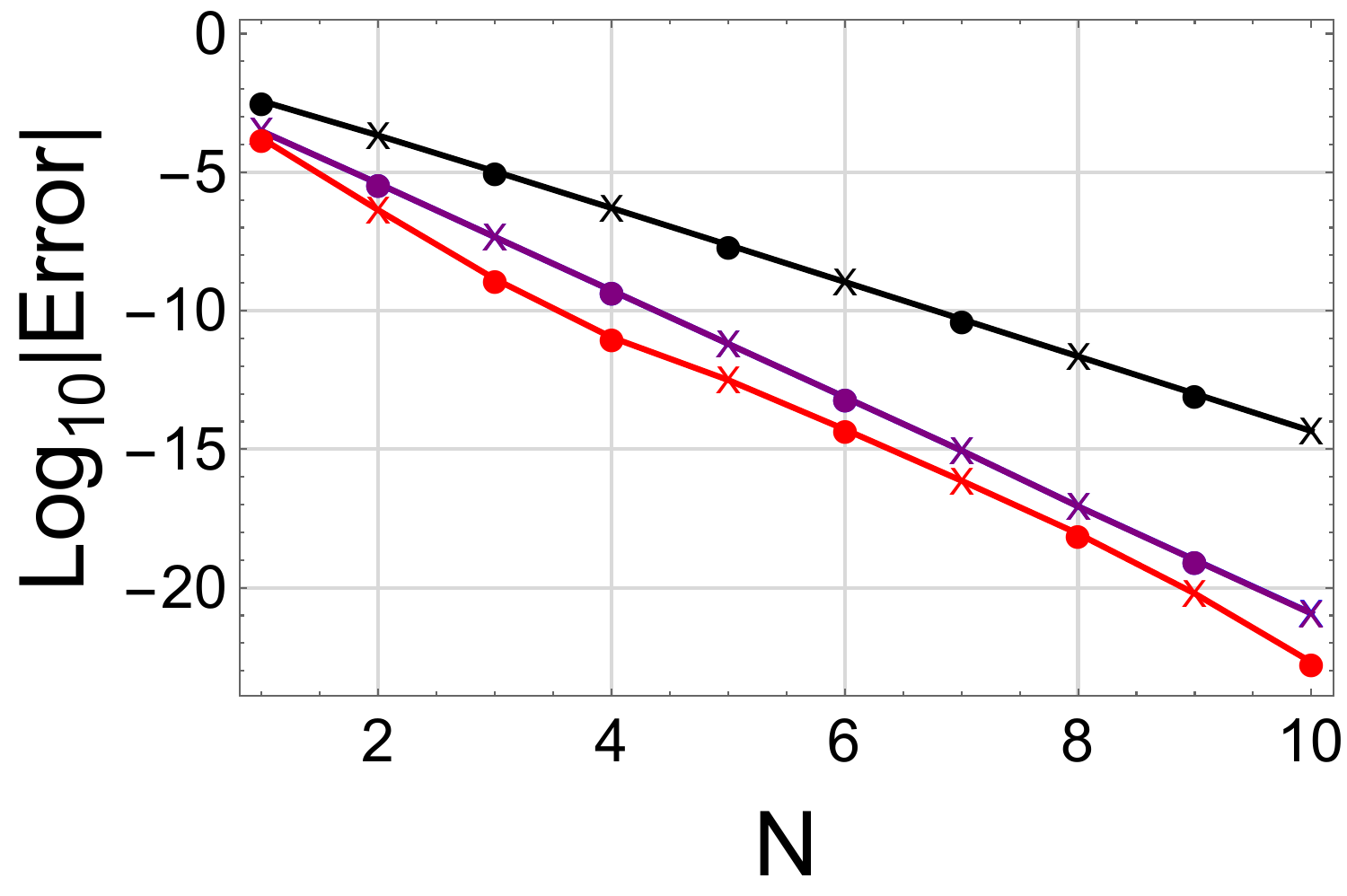}
\caption{Errors in approximations of sums: SWKB (black), CSWKB (purple), and HYP (red), using Table \ref{tab:SWKB.SC}, conventions as in Fig. \ref{fig:WKBEig}.}
\label{fig:SumTableErr}
\end{figure}

Table \ref{tab:SWKB.SC} shows results for various approximations to the sum of eigenvalues.  We plot the errors in Table \ref{tab:SWKB.SC} in Fig. \ref{fig:SumTableErr}.  The performance of SWKB for the sum is noticeably better than that of WKB for the eigenvalues.  Since $E(1)=\eps_0$, they are directly comparable for the lowest eigenvalue, and here SWKB is an order of magnitude better than WKB, because the optimal truncation occurs at $M$ = 3 for SWKB but at $M$ = 1 for WKB.  More remarkably, when subdominant corrections are included, the hyperasymptotic (HYP) approximation has an error about 100 times smaller than CWKB.  Thus, because it increases the optimal truncation order, the summation method produces much better results in even the most difficult case.  We can also compare the SWKB error for $N=10$ with the error of the WKB eigenvalue for $\n=9$:  The absolute error is about 20 times smaller (and the relative error is even better).  Finally, we mention that the sum of WKB eigenvalues can never compete with the asymptotic sums, as it is always dominated by its worst case, the ground-state value, as shown by the fourth column of the table. 

\begin{figure}[htb!]
\includegraphics[width=0.8\columnwidth]{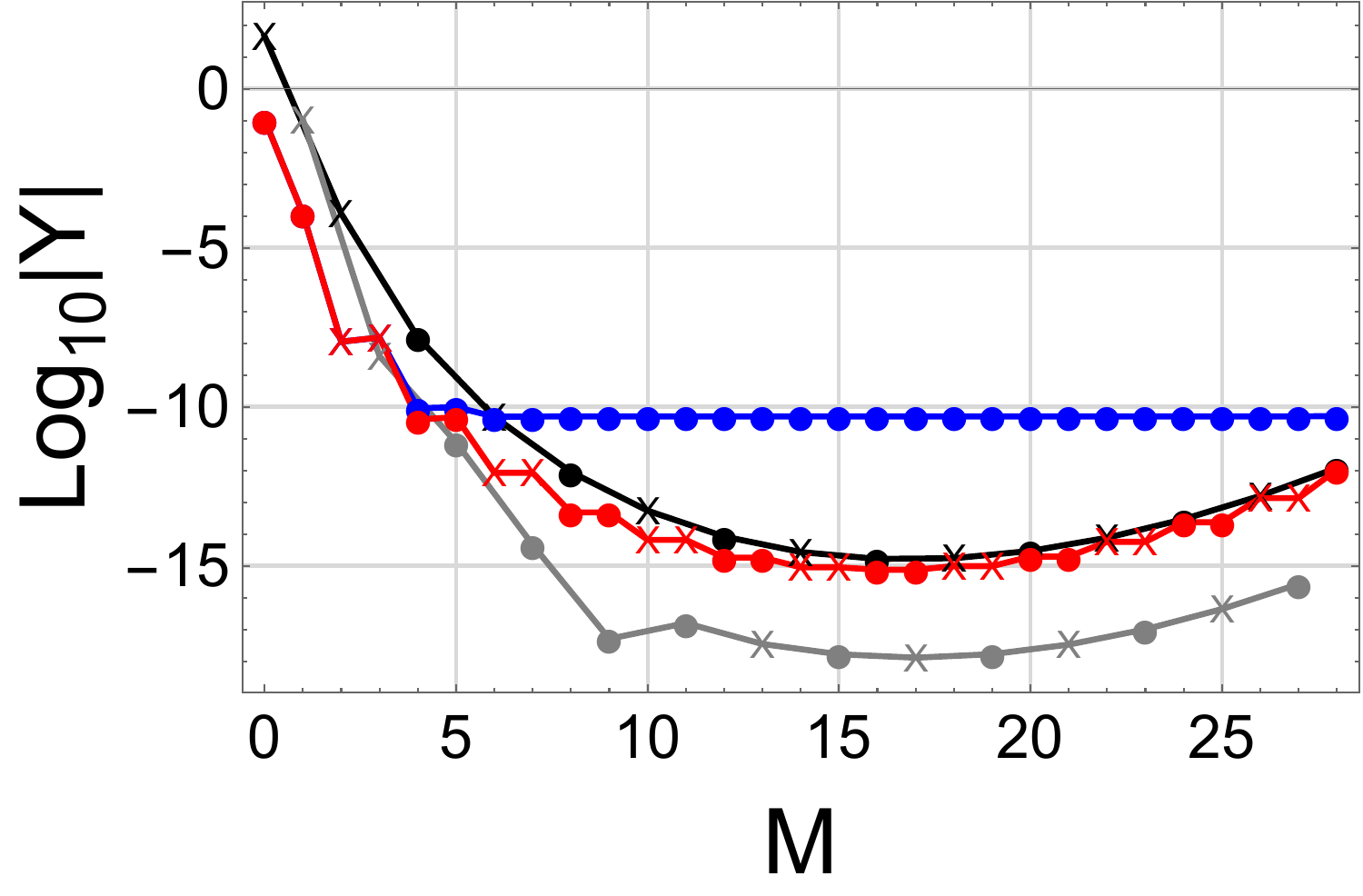}
\caption{Errors and additions of approximate sums for $N = 7$: SWKB (blue), CSWKB (red), even additions(black), odd additions (grey), with conventions as in Fig. \ref{fig:WKBEig}.}
\label{fig:EN7}
\end{figure}

Finally, we turn to the subdominant contributions to the asymptotics of the sums.  As $E(\infty) = 0$, so $E\SD(\infty) = 0$, yielding
\begin{equation}
\label{SDSum}
E\SD(N) = - \sum_{\n = N}^{\infty} \eps\SD_L(j + 1/2),
\end{equation}
where we use Eq. (\ref{eSD}) for $\eps\SD_L(z)$.  The results are shown for $N=7$ in Fig. \ref{fig:EN7}.  The corrections greatly improve over the SWKB results, as in the case of the eigenvalues (see Fig. \ref{fig:WKBvSC}).  Here the even SWKB additions closely follow the CSWKB error [Eq. (\ref{CSWKB})], while the odd additions are two orders of magnitude smaller.  We report the exact SD values and their asymptotic approximations (Eqs. (\ref{SDSum}) \& (\ref{AsymSumSD}) respectively), used in Table \ref{tab:SWKB.SC}, in Table \ref{tab:SumSD}.  We report the asymptotic values to more digits in Table S20.

\begin{table}[!ht]
$\begin{array}{|c|r|r|}
\hline
N & \mc{1}{c|}{\text{SD}} & \mc{1}{c|}{\text{Asym. SD}} \\
\hline
1 &  4.03447966131\x 10^{-03} &  4.03383052423\x 10^{-03} \\
2 & -2.17289733012\x 10^{-04} & -2.17287572833\x 10^{-04} \\
3 &  1.07361905207\x 10^{-05} &  1.07361692575\x 10^{-05} \\
4 & -5.10590048004\x 10^{-07} & -5.10589761517\x 10^{-07} \\
5 &  2.37706662505\x 10^{-08} &  2.37706614235\x 10^{-08} \\
6 & -1.09173453706\x 10^{-09} & -1.09173444207\x 10^{-09} \\
7 &  4.96721587271\x 10^{-11} &  4.96721566364\x 10^{-11} \\
8 & -2.24449297891\x 10^{-12} & -2.24449292884\x 10^{-12} \\
9 &  1.00887497460\x 10^{-13} &  1.00887496180\x 10^{-13} \\
10 & -4.51598363452\x 10^{-15} & -4.51598360000\x 10^{-15} \\
\hline
\end{array}$
\caption{Subdominant corrections to sums, $E\SD$, using Eq. (\ref{SDSum}) and its asymptotic approximation, Eq. (\ref{AsymSumSD}).}
\label{tab:SumSD}
\end{table}

Our last step is to turn the SD corrections to the sum into an asymptotic series.  We allow the order to vary in Eq. (\ref{SDSum}) and rewrite it as
\begin{equation}
\label{ExSD}
E\SD_M(N) = - \eps_M\SD(Z) \sum_{\n = 0}^{\infty} \frac{(-1)^\n}{e^{\pi \n}} \lf(1 + \frac{\n}{Z}\rg)^{1/3} \frac{H_M(\n + Z)}{H_M(Z)}.
\end{equation}
Expanding the above summand in powers of $Z$ and summing each term from $\n = 0$ to infinity (in this case we can evaluate the sums analytically but in a more general case we can still generate an asymptotic series using the Euler-Maclaurin formula, see Refs. \cite{B20b,H12} and Sec. 24.17 of Ref. \cite{DLMF}) yields
\begin{equation}
\label{AsymSumSD}
E\SD(N) = - \frac{\eps\SD(N)}{1 + e^{-\pi}} \sum_{n = 0}^{4} \frac{f_n}{Z^n},
\end{equation}
where we found that we only need the $f_n$ up to $n = 4$ to recover the same accuracy as Eq. (\ref{SDSum}).  As always $f_0 = 1$ and we give the other $f_n$ in Eq. (\ref{qCoefficients}).  Even for $\n=0$, these are always decreasing in magnitude, so we have not reached the least addition.

We can now define a corrected SWKB (CSWKB) approximation:
\begin{equation}
\label{CSWKB}
E\CSWKB_M(N) = E\SWKB_M(N) + E\SD(N),
\end{equation}
where we can use either Eq. (\ref{SDSum}) or (\ref{AsymSumSD}) for $E\SD(N)$.  Table \ref{tab:SWKB.SC} shows that both choices yield comparable accuracy.  Since the even additions determine the shape of the error, they determine the best truncation order.  We find the best results when we evaluate the SWKB contribution at the odd order immediately before the even least addition ( given in Table \ref{tab:SWKB.SC}).  This is true from $N = 8$ onward, but we do not have enough $c_n$ coefficients to see how far this trend goes.  Table S22 shows that the errors for orders near the least addition are similar.

Since the even additions are much larger in magnitude than the odd additions, we ignore the odd additions and note that the remaining even additions alternate in sign.  We can therefore apply hyperasymptotics in exactly the same fashion as for the linear half-well \cite{BB20}, adding half of the optimal SWKB term plus half of the term two orders larger.  Then we add this quantity to the asymptotic approximation to $E\SD(N)$ in Eq. (\ref{AsymSumSD}).  The end result is
\begin{equation}
\label{HYP}
E^{\text{HYP}}_M(N) = \frac{E\SWKB_M(N) + E\SWKB_{M+2}(N)}{2} + E\SD(N).
\end{equation} 
This hyperasymptotic approximation improves results by up to two orders of magnitude over CSWKB, as shown in Table \ref{tab:SWKB.SC}.  Table S22 shows the hyperasymptotic energies are more sensitive to changes in order than the CSWKB energies.

\input{TableWKBvSWKBEigenvalues}

\begin{figure}[htb!]
\includegraphics[width=0.8\columnwidth]{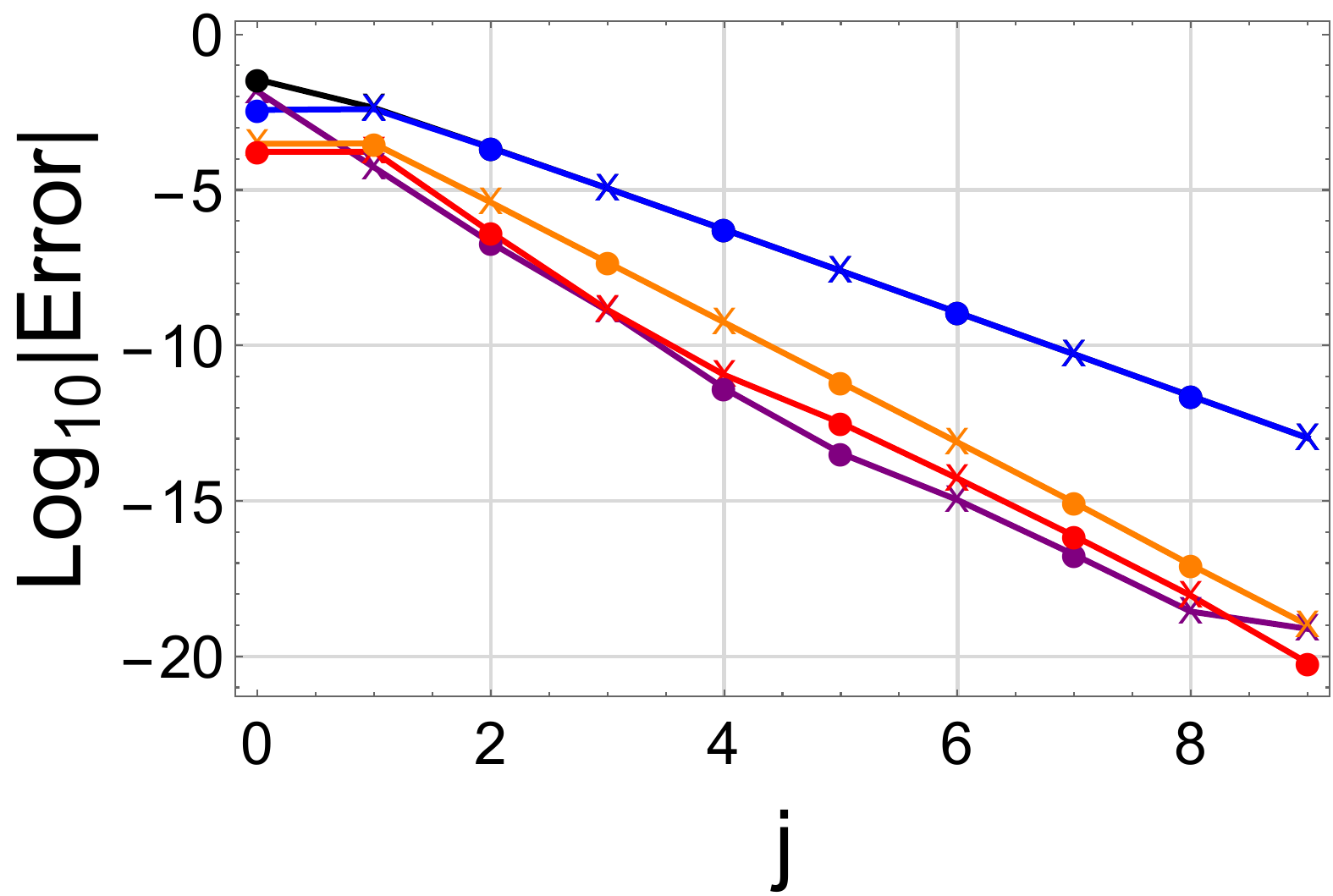}
\caption{Errors in individual eigenvalues, including differences of approximate sums: WKB (black), SWKB (blue), CWKB (purple), CSWKB (orange), and HYP (red), using Table \ref{tab:CompWKB.BB.Eig} and the conventions of Fig. \ref{fig:WKBEig}.}
\label{fig:SWKB_WKB_COMP}
\end{figure}

Next we use the various approximations for sums developed here to find new approximations for the eigenvalues, by taking differences of sums:
\begin{equation}
\label{EigValFromSum}
\eps_\n = E(\n+1) - E(\n).
\end{equation}
We compare our sum approximations with the eigenvalue approximations of Sec. \ref{sec:EigVal}  in Table \ref{tab:CompWKB.BB.Eig} and plot the errors in Fig. \ref{fig:SWKB_WKB_COMP}.  As mentioned above, the SWKB result is an order of magnitude better for $j=0$, but otherwise their errors are almost identical.  However, when subdominant corrections are included, the CWKB results generally cannot be beaten, even with hyperasymptotic corrections to the sums.  Equation (\ref{SumForm}) also yields a simple approximation for $E(N) + \eps_N/2$.  In Appendix \ref{sec:NHalf}, we check that such midway sums do not yield sums of higher accuracy than those presented here.

\sec{Conclusion and Relevance to DFT}
\label{sec:Con}

In the first half of this paper, we analyzed the semiclassical expansion of the quartic oscillator eigenvalues.  Although the formal structure of this expansion has been known since the late 1970s, we have provided a thorough analysis of the performance of the semiclassical expansion and reported the coefficients of the relevant asymptotic series to many orders and many digits.  Moreover, in Eqs. (\ref{ActionCoeff}), (\ref{Form.a}), and (\ref{Rec.b}), we give the explicit formulas to construct the implicit and explicit WKB series to any order.  This should prove useful for chemists, mathematicians, or physicists developing methods in the many different contexts in which the quartic oscillator is studied.

But our primary results occur in the second half, where we report the first application of the asymptotics of sums to a semiclassical expansion with subdominant contributions (beyond the WKB series), due to the presence of complex turning points.  The method of asymptotics of sums, first demonstrated in Ref. \cite{BB20}, can be easily adapted (at least in this case) to produce subdominant corrections to the sums of eigenvalues.  When these are included, the performance for sums is better than the original series for the individual eigenvalues, often by more than an order of magnitude.  We can employ hyperasymptotics to improve performance even further, yielding errors less than $10^{-22}$ for the sum of the first 10 terms.  Unsurprisingly, taking differences of the asymptotic sums does not in general yield better results than the original series, except for the ground state.

Local density functionals become relatively exact in a well-defined semiclassical limit \cite{OB21}, but the leading corrections to these local functionals in this limit are unknown.  Modern GGAs are approximations to these corrections.  We do not have a general procedure for deriving these corrections for the unknown XC energy of Kohn-Sham (KS) \cite{KS65} DFT, or even for the non-interacting kinetic energy in three dimensions of orbital-free DFT.  But the asymptotics of sums of eigenvalues is directly related to orbital-free DFT in one dimension because one considers the levels as filled by same-spin fermions.  Modern KS-DFT calculations iteratively solve for non-interacting electrons in a density-dependent potential, with some given approximation for exchange-correlation effects \cite{WB13}.  The expressions developed here directly yield the sum of the energies of the lowest $N$ orbitals of non-interacting fermions.   Thus, following the Aufbau principle, this would yield the sum of the occupied KS levels in a single iteration. If accurate and general analogs could be found for three-dimensional systems, along with similar analogs to yield three-dimensional densities \cite{RLCE15,RB17,RB18}, the need to solve the KS equations (typically the biggest bottleneck in such calculations) would be by-passed, allowing KS-DFT to be applied to much larger systems than is currently possible.  Using the spin-scaling relation for the kinetic energy \cite{ABC}, and going from potential functionals to density functionals \cite{CLEB11,CGB13}, one can relate the asymptotic expansion for the sum of eigenvalues to the expansion of the non-interacting kinetic energy around the semiclassical limit.  Thus, the zero-order contribution to that sum is a simple local functional of the potential.  In the full semiclassical limit, this result is identical to that of non-interacting Thomas-Fermi (TF) theory, i.e., the result given by a local density approximation to the non-interacting kinetic energy \cite{MP56}.  Likewise, the leading asymptotic correction to the sum yields the leading correction to the local approximation, and is determined both by the derivatives of the potential (up to the second) and by the Maslov index for the problem.   Our results here show that, in principle, far more accurate results can be extracted from these semiclassical expansions than is currently possible with density functional approximations.  Inclusion of even the exact leading correction ($M=1$ here) can greatly improve results \cite{B20}.

The work in this paper produces highly accurate results for a specific potential, but the general methodology has been outlined in Refs. \cite{B20b,BB20,OB21} for an arbitrary (but sufficiently smooth) potential.  The procedures given there yield results as functionals of the potential.  In general, such potential-functionals can be used directly on given problems (and could be combined with density-functionals for exchange-correlation).  But since the world of electronic structure calculations is more familiar with density functionals, Ref. \cite{B20b} showed how one converts such potential functionals to density functionals.  This is a non-trivial process to be carried out order-by-order, and the higher-order corrections yield density functionals that are very different from any existing approximations.

One could apply the general form of this methodology to one-dimensional interacting systems, which can be accurately and efficiently solved using the density-matrix renormalization group \cite{W92} or quantum Monte Carlo methods \cite{PZTA18}, and have been used as a mimic of realistic three-dimensional DFT calculations \cite{SWWB12,BSWB15,SWWB12}.  Related semiclassical developments for the first-order density matrix were used to accurately approximate exchange energies \cite{ECPG15}.  The general form of the methods discussed in Refs. \cite{B20b,BB20,OB21} can be immediately applied to the one-dimensional simulation used in Ref. \cite{SWWB12}.   However, such systems are sufficiently different from realistic systems that we see little value in performing such calculations.  Our over-arching goal is to develop methods for three-dimensional systems.   Moreover, the value of by-passing the Kohn-Sham equations is far greater in three dimensions than in one.

The subdominant corrections appearing in the current work are largest for $N = 1$ (one particle in the lowest eigenvalue), where they would be non-negligible in an electronic structure calculation.  The present work shows that such terms can be handled in terms of asymptotics of sums.  Of course, this would also require the further step of finding a general expression for their form and coefficients, not just the value for a quartic oscillator, but that step is beyond the scope of the current work.

\sec{Acknowledgments}
We thank the NSF (CHE-2154371) for funding and Micheal Berry for useful discussions.  KB will always be grateful to Michael Berry for a rather late-in-life whirlwind introduction to semiclassical methods and asymptotic expansions.  He expects to use the wisdom imparted to greatly improve insight into density functional approximations, and also hopes to improve both their speed and their accuracy.

\appendix
\input{DeriveA}
\input{Coeffs}
\input{AppNHalf}
\clearpage
\sec{Table of Constants}
\label{sec:TabConst}
\input{ConstantsTab}

\end{document}

%% file: TableEigValApprox.tex
\begin{table*}[!ht]
$\begin{array}{|c|c|c|r|r|r|r|r|}
\hline
\mc{4}{|c|}{} & \mc{4}{c|}{\text{Error}} \\
\hline
\n & M & L & \mc{1}{c|}{\eps}  & \mc{1}{c|}{\text{WKB}} & \mc{1}{c|}{\text{BCWKB}} & \mc{1}{c|}{\eps\WKB_M + \D \eps_L} & \mc{1}{c|}{\text{CWKB}}\\
\hline
 0 &  1 & 1  &  0.5301810 & -3.5270398\x 10^{-02} &  1.8285750\x 10^{-02} &  1.7091440\x 10^{-02} &  1.4571908\x 10^{-02} \\
 1 &  2 & 3  &  1.8998365 &  4.3044942\x 10^{-03} &  5.6300300\x 10^{-05} &  5.2731082\x 10^{-05} &  5.2724811\x 10^{-05} \\
 2 &  6 & 6  &  3.7278490 & -2.2822571\x 10^{-04} & -1.9645769\x 10^{-07} & -1.9824144\x 10^{-07} & -1.9979145\x 10^{-07} \\
 3 &  8 & 8  &  5.8223728 &  1.1248072\x 10^{-05} &  1.2933375\x 10^{-09} &  1.2905883\x 10^{-09} &  1.2914117\x 10^{-09} \\
 4 & 10 & 11 &  8.1309130 & -5.3436485\x 10^{-07} & -4.1292807\x 10^{-12} & -4.1333145\x 10^{-12} & -4.1333154\x 10^{-12} \\
 5 & 12 & 13 & 10.6191865 &  2.4862367\x 10^{-08} & -3.3798630\x 10^{-14} & -3.3805755\x 10^{-14} & -3.3805755\x 10^{-14} \\
 6 & 14 & 15 & 13.2642356 & -1.1414056\x 10^{-09} &  1.0688951\x 10^{-15} &  1.0688829\x 10^{-15} &  1.0688829\x 10^{-15} \\
 7 & 16 & 17 & 16.0492989 &  5.1916633\x 10^{-11} & -1.8636440\x 10^{-17} & -1.8636461\x 10^{-17} & -1.8636461\x 10^{-17} \\
 8 & 18 & 19 & 18.9615005 & -2.3453802\x 10^{-12} &  2.7529489\x 10^{-19} &  2.7529485\x 10^{-19} &  2.7529485\x 10^{-19} \\
 9 & 21 & 21 & 21.9905790 &  1.0540356\x 10^{-13} &  7.6509652\x 10^{-20} &  7.6509652\x 10^{-20} &  7.6509652\x 10^{-20} \\
10 & 23 & 23 & 25.1281273 & -4.7174532\x 10^{-15} & -8.9559882\x 10^{-22} & -8.9559882\x 10^{-22} & -8.9559882\x 10^{-22} \\
\hline
\end{array}$
\caption{Errors in WKB series, including various approximations for subdominant contributions: WKB [Eq. (\ref{WKBExpl})], BCWKB [Eq. (\ref{CWKB})], $\eps\WKB_M + \D \eps_L$ [Eq. (\ref{SC1})], and CWKB [Eqs. (\ref{eSD}) \& (\ref{DefCWKB})].  $M$ is the order at which we truncate, and is one less than the order of least addition. $L$ is the order of least addition for Eq. (\ref{SC1}).}
\label{tab:SCError}
\end{table*}

%% file: TableSumApprox.tex
\begin{table*}
$\begin{array}{|c|c|r|c|r|r|r|r|}
\hline
\mc{3}{|c|}{} & \mc{5}{c|}{\text{Error}} \\
\hline
N  & M & \mc{1}{c|}{\text{Exact}} & \Sig\text{WKB} & \mc{1}{c|}{\text{SWKB}} & \mc{1}{c|}{\text{CSWKB}} & \mc{1}{c|}{\text{Asym. CSWKB}} & \mc{1}{c|}{\text{HYP}} \\
\hline
 1 &  3 &   0.53018 & -3.52704\x 10^{-2} & -3.72955\x 10^{-03} &  3.04929\x 10^{-04} &  3.04280\x 10^{-04} & -1.68851\x 10^{-04} \\
 2 &  5 &   2.43002 & -3.09659\x 10^{-2} &  2.13376\x 10^{-04} & -3.91377\x 10^{-06} & -3.91161\x 10^{-06} &  4.31969\x 10^{-07} \\
 3 &  7 &   6.15787 & -3.11941\x 10^{-2} & -1.06904\x 10^{-05} &  4.58054\x 10^{-08} &  4.57841\x 10^{-08} & -1.40070\x 10^{-09} \\
 4 &  9 &  11.98024 & -3.11829\x 10^{-2} &  5.10050\x 10^{-07} & -5.40474\x 10^{-10} & -5.40187\x 10^{-10} & -1.09301\x 10^{-11} \\
 5 & 11 &  20.11115 & -3.11834\x 10^{-2} & -2.37643\x 10^{-08} &  6.38318\x 10^{-12} &  6.37835\x 10^{-12} &  3.15656\x 10^{-13} \\
 6 & 13 &  30.73034 & -3.11834\x 10^{-2} &  1.09166\x 10^{-09} & -7.54597\x 10^{-14} & -7.53647\x 10^{-14} & -5.18497\x 10^{-15} \\
 7 & 15 &  43.99457 & -3.11834\x 10^{-2} & -4.96713\x 10^{-11} &  8.93476\x 10^{-16} &  8.91385\x 10^{-16} &  7.31437\x 10^{-17} \\
 8 & 19 &  60.04387 & -3.11834\x 10^{-2} &  2.24450\x 10^{-12} &  8.58161\x 10^{-18} &  8.63167\x 10^{-18} & -8.89917\x 10^{-19} \\
 9 & 21 &  79.00537 & -3.11834\x 10^{-2} & -1.00888\x 10^{-13} & -9.97495\x 10^{-20} & -1.01030\x 10^{-19} &  6.14869\x 10^{-21} \\
10 & 23 & 100.99595 & -3.11834\x 10^{-2} &  4.51598\x 10^{-15} &  1.16502\x 10^{-21} &  1.19954\x 10^{-21} & -2.03952\x 10^{-23} \\
\hline
\end{array}$
\caption{Errors in various approximations for sums: $\Sig$WKB (sum of WKB eigenvalues from Table \ref{tab:SCError}),  SWKB from Eq. (\ref{PQO.BB}), CSWKB from Eq. (\ref{CSWKB}) with subdominant corrections of Eq. (\ref{SDSum}), Asym. CSWKB using Eq. (\ref{AsymSumSD}), and the hyperasymptotic (HYP) approximation from Eq. (\ref{HYP}).}
\label{tab:SWKB.SC}
\end{table*}

%% file: TableWKBvSWKBEigenvalues.tex
\begin{table*}
$\begin{array}{|c|r|r|r|r|r|r|}
\hline
\mc{2}{|c|}{} & \mc{5}{c|}{\text{Error}}\\
\hline
j & \mc{1}{c|}{\text{Exact}} & \mc{1}{c|}{\text{WKB}} & \mc{1}{c|}{\text{SWKB}} & \mc{1}{c|}{\text{CWKB}} & \mc{1}{c|}{\text{CSWKB}} & \mc{1}{c|}{\text{HYP}} \\
\hline
0 &  0.53018 & -3.52704\x 10^{-02} & -3.72955\x 10^{-03} &  1.45719\x 10^{-02} &  3.04280\x 10^{-04} & -1.68851\x 10^{-04} \\
1 &  1.89984 &  4.30449\x 10^{-03} &  3.94293\x 10^{-03} &  5.27248\x 10^{-05} & -3.08191\x 10^{-04} &  1.69283\x 10^{-04} \\
2 &  3.72785 & -2.28226\x 10^{-04} & -2.24066\x 10^{-04} & -1.99791\x 10^{-07} &  3.95739\x 10^{-06} & -4.33369\x 10^{-07} \\
3 &  5.82237 &  1.12481\x 10^{-05} &  1.12004\x 10^{-05} &  1.29141\x 10^{-09} & -4.63243\x 10^{-08} &  1.38977\x 10^{-09} \\
4 &  8.13091 & -5.34365\x 10^{-07} & -5.33814\x 10^{-07} & -4.13332\x 10^{-12} &  5.46566\x 10^{-10} &  1.12457\x 10^{-11} \\
5 & 10.61919 &  2.48624\x 10^{-08} &  2.48559\x 10^{-08} & -3.38058\x 10^{-14} & -6.45372\x 10^{-12} & -3.20841\x 10^{-13} \\
6 & 13.26424 & -1.14141\x 10^{-09} & -1.14133\x 10^{-09} &  1.06888\x 10^{-15} &  7.62561\x 10^{-14} &  5.25811\x 10^{-15} \\
7 & 16.04930 &  5.19166\x 10^{-11} &  5.19158\x 10^{-11} & -1.86365\x 10^{-17} & -8.82753\x 10^{-16} & -7.40337\x 10^{-17} \\
8 & 18.96150 & -2.34538\x 10^{-12} & -2.34539\x 10^{-12} &  2.75295\x 10^{-19} & -8.73270\x 10^{-18} &  8.96066\x 10^{-19} \\
9 & 21.99058 &  1.05404\x 10^{-13} &  1.05404\x 10^{-13} &  7.65097\x 10^{-20} &  1.02230\x 10^{-19} & -6.16909\x 10^{-21} \\
\hline
\end{array}$
\caption{Errors in approximate eigenvalues, including differences of approximate sums:
WKB and CWKB from Table \ref{tab:SCError}, SWKB, CSWKB, and hyperasymptotic CSWKB of Table \ref{tab:SWKB.SC} plugged into Eq. (\ref{EigValFromSum}), with the asymptotic subdominant corrections of Eq. (\ref{AsymSumSD}).}
\label{tab:CompWKB.BB.Eig}
\end{table*}

%% file: DeriveA.tex
\sec{Deriving the Implicit WKB Series}
\label{sec:DeriveImpWKB}
Following Bender \& Orszag \cite{BO99} we write the WKB wavefunction as $\psi(x) = \exp[i S(x)]$ and expand the action in a semiclassical series $S(x) = S_0(x) + S_1(x) + \cdots$.  For the pure quartic oscillator
\begin{equation}
\label{DerivAction}
S_n'(x)  = (-1)^{\floor{(n+1)/2}}i^{n+1} \sum_{m = 0}^{\floor{3n/4}} \frac{k_n^{(m)} x^{3n-4m}}{[p(x)]^{3n-2m-1}},
\end{equation}
where $p(x) = \sqrt{2\eps - x^4}$ is the classical momentum and the prime indicates differentiation, and $k_0^{(0)} = k_1^{(0)} = 1$ \cite{BO99}.  The higher order ($n > 1$) action coefficients are determined by
\begin{equation}
\label{HighOrderAction}
2 S_0' S_n' + S_{n-1}'' + \sum_{m = 1}^{n - 1} S_m'S_{n-m}' = 0.
\end{equation}
This recursion relation, derived in Ref. \cite{BO99}, yields
\begin{multline}
\label{ActionCoeff}
k_n^{(m)} = (-1)^n \bigg[(3n-2m-4) k_{n-1}^{(m)} +\\
\frac{3n-4m+1}{2} k_{n-1}^{(m-1)}\bigg] - \half \sum_{l = 1}^{n-1} (-1)^{l(n+1)} \sum_{r = 0}^{m} k_l^{(r)} k_{n-l}^{(m-r)},\\
\end{multline}
for $n > 1$.  To use Eq. (\ref{ActionCoeff}), we set all coefficients $k_n^{(m)}$ with $m > \floor{3n/4}$ and $m < 0$ to 0.  In general each $k_n^{(m)}$ is a function of the set $\{k_{n'}^{(m')}\}$ with $m' \leq m$ and $n' < n$.  We give the $k_n^{(m)}$ up to $n = 21$ in Table S4.  The eigenvalues are determined only by the $S_n(x)$ with even $n$ so we drop the odd $n$ values and integrate Eq. (\ref{DerivAction}):
\begin{equation}
S_{2n}(\eps) = 2 i \sum_{m = 0}^{\floor{3n/2}} k_{2n}^{(m)} \int_0^{x_0} dx\, \frac{x^{6n-4m}}{[p(x)]^{6n - 2m - 1}},
\end{equation}
where $x_0 = x_0(\eps)$ is the classical turning point defined by $p[x_0(\eps)] = 0$.  Formally there are divergences at $x_0$ but we remove them by integrating under the integral sign with respect to $\eps$ \cite{OB21b}.  We know from Ref. \cite{BO99} that 
\begin{equation}
\label{aRelateToAction}
S_{2n}(\eps) = \frac{i\, a_n \g^2 (2\eps)^{3/4-3n/2}}{3\sqrt{2\pi}},
\end{equation}
so after some algebra we get
\begin{equation}
\label{Form.a}
a_n =  \frac{3}{\g^2} \sqrt{\frac{\pi}{2}} \sum_{m = 0}^{\floor{3n/2}} k_{2n}^{(m)} B \left( m - 3n + \frac{3}{2},\frac{3n}{2} - m + \frac{1}{4} \right),
\end{equation}
where $B(x,y)$ is the incomplete Euler beta function $\G(x)\G(y)/\G(x+y)$ (Sec. 8.17 of Ref. \cite{DLMF}).  Equation (\ref{Form.a}) was found by Voros, Balian, and Parisi in Ref. \cite{BPV79}.  The first three nontrivial ($n = 1,2,3$) $a_n$ are
\begin{equation}
- \frac{3\pi^2}{2\g^4}, \qquad  \frac{11}{512}, \qquad  \frac{4697 \pi^2}{5120 \g^4}.
\end{equation}

%% file: Coeffs.tex
\sec{Analytic Expressions for Coefficients}
\label{sec:AnalyticStructureCoeff}

We generate the $b_n$ in Eq. (\ref{WKBExpl}) using the $a_n$ and lower order $b_n$:
\begin{equation}
\label{Rec.b}
b_n = \sum_{m = 0}^{n} a_m \sum_{r = 0}^{n-m} G_{n m r} \mathscr{P}_r^{(n-m)}(\mbf{b}), \qquad n \geq 1,
\end{equation}
where $\mathscr{P}_0^{(n)} = \d_{n,0}$ and
\begin{equation}
\mathscr{P}_m^{(n)}(\mbf{b}) = \sum_{\substack{k_1,k_2,...,k_m = 1 \\ k_1 + k_2 + ... + k_m = n}}^{n + 1 - m} b_{k_1}b_{k_2}...b_{k_m},
\end{equation}
e.g. $\mathscr{P}_3^{(5)}(\mbf{b}) = 3b_1 b_2^2 + 3b_1^2 b_3$, and $G_{n 0 1} = 0$ and
\begin{equation}
G_{n m r} = - \frac{4}{3(2\a)^{3m/2}} \binom{3/4-3m/2}{r}.
\end{equation}

The coefficients we have derived in this paper ($b_n$, $c_n$, $d_n$, $h_n$, $k_n$, $f_n$) can each be expressed in terms of the $a_n$ coefficients.  In general each set of coefficients is composed of more fundamental coefficients.  Here we give both the relation to simpler coefficients and explicit exact formulas.  The supplemental info contains all tables needed to reconstruct these, as well as numerical values to 40 digits.  We remind the reader that our convention is to set the $n = 0$ coefficient to 1.

Eliminating lower order $b_n$'s from Eq. (\ref{Rec.b}) yields the $b_n$ in terms of the $a_n$ alone:
\begin{equation}
\label{b.in.terms.of.a}
b_n = - \sum_{m = 0}^{n - 1} L_{n m} \mathscr{P}^{(n)}_{n-m}(\mbf{a}), ~~ n \geq 1,
\end{equation}
where we give the $L_{n m}$ in Table S6.  Inserting the $a_n$ from Eq. (\ref{Form.a}) yields exact values for the $b_n$ explicitly given in Table S7.  The first 3 nontrivial $b_n$ are
\begin{equation}
\frac{1}{9\pi}, ~~ -\frac{1}{648\pi^2} \lf(5 + \frac{11\g^8}{192\pi^4}\rg), ~~ \frac{11}{2916\pi^3} \lf(\frac{1}{3} - \frac{31\g^8}{640\pi^4}\rg).
\end{equation}

Similarly, we can express the $d_n$ in terms of the $a_n$ and $b_n$:
\begin{equation}
\label{dAB}
d_n = \sum_{m = 0}^{n} a_{n-m} \sum_{r = 0}^{m}  F_{n m r} \mathscr{P}^{(m)}_r(\mbf{b}),
\end{equation}
where we give the $F_{n m r}$ in Table S9.  Inserting the $a_n$ from Eq. (\ref{Form.a}) and the $b_n$ from Eq. (\ref{b.in.terms.of.a}) yields explicit expressions for the $d_n$ which we give in Table S10.  The first 3 nontrivial $d_n$ are
\begin{equation}
\frac{1}{6\pi}, \quad  - \frac{1}{72 \pi^2}, \qquad \frac{1}{432\pi^3}\lf(1 - \frac{1133 \g^8}{8640\pi^4}\rg).
\end{equation}

For our next set, it is best to represent the $h_n$ directly in terms of the $a_n$:
\begin{equation}
\label{h.A}
h_n = \sum_{m = 0}^{\floor{n/2}} P_{n m} (\mbf{a}),
\end{equation}
where the $P_{n m} (\mbf{a})$ are polynomials of $a_n$, given in Table S12, for which we have not been able to find a general pattern.  We give the $h_n$ themselves in Table S13.  The first 3 nontrivial $h_n$ are
\begin{equation}
\label{First3h}
- \frac{1}{6}, \qquad \frac{1}{72}\lf(1 - \frac{4}{\pi}\rg), \qquad \frac{1}{216} \lf(\frac{5}{\pi} - \frac{1}{6}\rg).
\end{equation}

We give the $c_n$ in terms of the $b_n$ in Eqs. (\ref{Form.C}) \& (\ref{Const.C}).  We give the exact forms of the $c_n$ up to $n=10$ in Table S15.  The first 3 nontrivial $c_n$ are
\begin{multline}
\frac{7}{9\pi} \lf(1 - \frac{\pi}{6}\rg), \qquad \frac{7}{324} \lf(\frac{1}{3\pi} + \frac{1}{2\pi^2} - \frac{7}{180} + \frac{11\g^8}{1920\pi^6}\rg ),\\
\frac{7}{2916} \lf[\frac{31}{756} - \frac{7}{18\pi} - \frac{5}{6\pi^2} - \frac{1}{3\pi^3} - \frac{\g^8}{128\pi^6}\lf(\frac{11}{9} - \frac{31}{5\pi}\rg)\rg].\\
\end{multline}

We give the $f_n$ in terms of the $h_n$ using $f_n = k_n/(1 + e^\pi)$ where
\begin{align}
\label{qCoefficients}
\begin{split}
k_1 &= - 1/3,\\
k_2 &= \frac{\tanh(\pi/2)}{9} + h_1,\\
k_3 &= - \frac{5(1 - 4 e^\pi + e^{2\pi})}{81(1 + e^\pi)^2} - \frac{2h_1}{3} \tanh\lf(\frac{\pi}{2}\rg) - h_1^2 + 2h_2,\\
k_4 &=  \frac{20 e^\pi (\cosh \pi - 5)}{243 (1 + e^\pi)^2} \tanh\lf(\frac{\pi}{2}\rg) + \frac{5(1 - 4 e^\pi + e^{2\pi})h_1}{9(1 + e^\pi)^2}\\
& + \frac{2h_1^2 - 7h_2}{3} \tanh\lf(\frac{\pi}{2}\rg) + h_1^3 + 3(h_3 - h_1 h_2).\\
\end{split}
\end{align}
And the first three nontrivial $h_n$ are given exactly in Eq. (\ref{First3h}).

%% file: AppNHalf.tex
\sec{Midway sums}
\label{sec:NHalf}

\input{TableGApprox}

An observant reader will have noticed that  Eq. (\ref{GExpan}) yields a simpler expression for the following midway sum:
\begin{equation}
G(N)= E(N) + \frac{\eps_N}{2}.
\end{equation}
By rearranging Eq. (\ref{GExpan}) we find an SWKB approximation for $G(N)$:
\begin{equation}
\label{GSWKB}
G\SWKB_M(N)=\frac{3}{7} \a Z^{7/3} Q_M(Z^{2}).
\end{equation}
While this is not the sum of primary interest, we explore here if this might yield more accurate approximations for $E(N)$.  The first three nontrivial $q_n$ are
\begin{multline}
\frac{7}{9} \lf(\frac{1}{3} + \frac{1}{\pi}\rg), \qquad \frac{7}{162} \lf( \frac{1}{45} - \frac{1}{3\pi} + \frac{1}{4\pi^2}\rg) + \frac{77 \g^8}{622080 \pi^6},\\
- \frac{1}{2187} \lf(\frac{2}{9} - \frac{7}{3\pi} - \frac{35}{4\pi^2} + \frac{7}{4\pi^3}\rg) + \frac{7 \g^8}{186624 \pi^6} \lf( \frac{11}{9} + \frac{31}{10 \pi}\rg).\\
\end{multline}
We give the $q_n$ exactly up to $n = 10$ in Table S17 and  numerically in Table S16.  Equation (\ref{q_of_b}) gives the $q_n$ in terms of the $b_n$.

It is easy to extract the SD contribution to $G$ from that of $E$:
\begin{equation}
G\SD(N) = \frac{E\SD(N + 1) + E\SD(N)}{2},
\end{equation}
where we use Eq. (\ref{AsymSumSD}) for $E\SD$.  Thus 
\begin{equation}
G\CSWKB_M(N) = G\SWKB_M(N) + G\SD(N). 
\end{equation}

\begin{figure}[htb!]
\includegraphics[width=0.8\columnwidth]{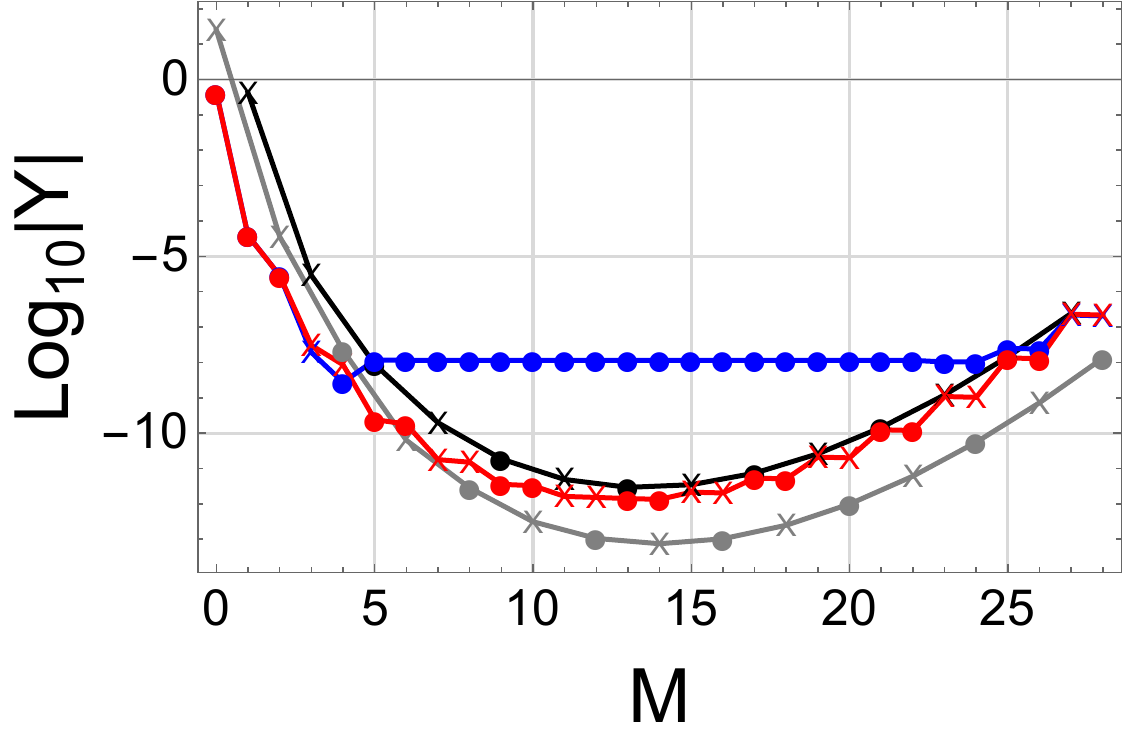}
\caption{The even (gray) and odd (black) SWKB additions with the SWKB (blue) and CSWKB (red) errors for $N = 5$.}
\label{fig:GN5}
\end{figure}

\begin{figure}[htb!]
\includegraphics[width=0.8\columnwidth]{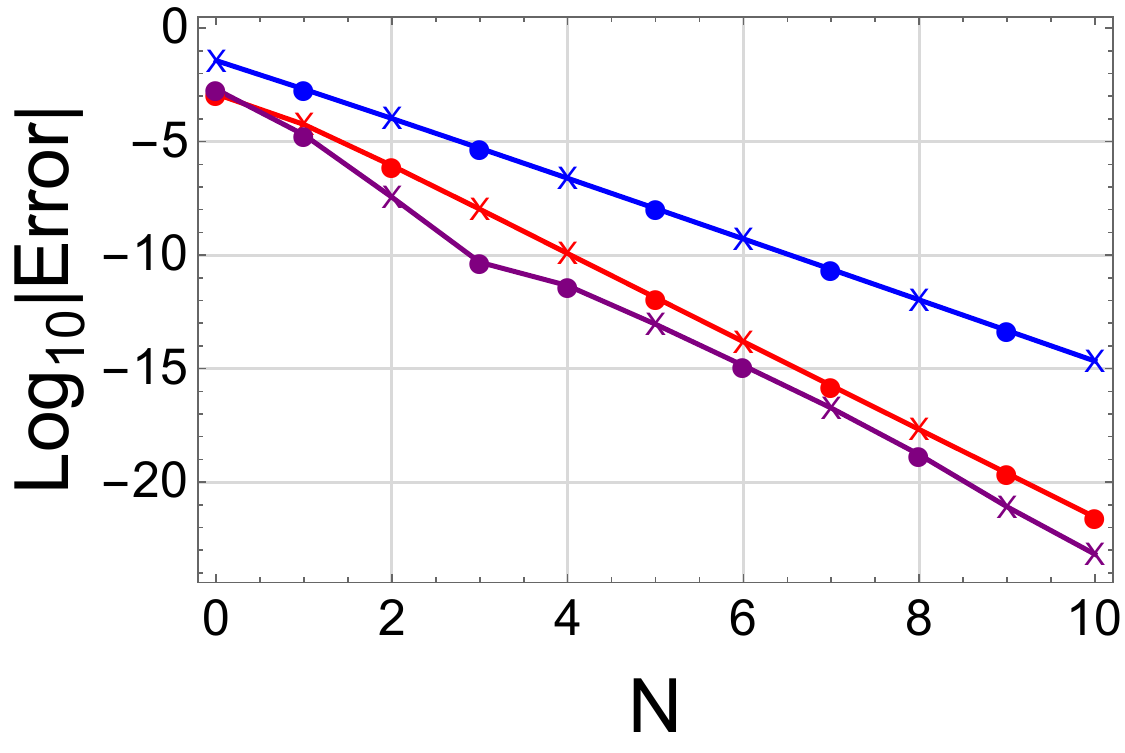}
\caption{The errors in Table \ref{tab:GApprox}: SWKB (blue), CSWKB (red), and HYP (purple).}
\label{fig:GTableErr}
\end{figure}

First, we consider how accurately we can approximate $G(N)$.  In Fig. \ref{fig:GN5} we show the SWKB additions and compare the SWKB and CSWKB errors for $N = 5$.  As always, the SWKB errors plateau but the SD corrections increase the accuracy by many orders of magnitude.  Since the even SWKB additions are much smaller in magnitude than the odd additions, a hyperasymptotic approximation to $G$ is
\begin{equation}
\label{hyperG}
G^{\text{HYP}}_M(N) = \frac{G\SWKB_M(N) + G\SWKB_{M+2}(N)}{2}  + G\SD(N).
\end{equation}
We compare the errors of the SWKB, CSWKB, and hyperasymptotic approximations to $G(N)$ in Table \ref{tab:GApprox}, while  Fig. \ref{fig:GTableErr} shows the errors.  Comparing Tables \ref{tab:SWKB.SC} and \ref{tab:GApprox} shows that the CSWKB and hyperasymptotic approximations are more accurate for $G(N)$ than for $E(N)$.

It would be nice if we could make use of this accuracy to turn the $G(N)$ approximation into an $E(N)$ approximation.  We can do this by subtracting half of the Nth eigenvalue from our $G(N)$ approximation.  Table \ref{tab:CompWKB.BB.Eig} shows that CWKB is our most accurate eigenvalue approximation.  We show in Table \ref{tab:CompGandE} that combining this with our hyperasymptotic $G(N)$ approximation gives a more accurate $E(N)$ approximation than the hyperasymptotic $E(N)$ (our best $E(N)$ approximation in Table \ref{tab:SWKB.SC}).  Thus we need to mix the $G(N)$ approximation with an eigenvalue approximation to generate an $E(N)$ approximation.  We prefer to work with pure $E(N)$ approximations instead.  Thus we relegate our discussion of the $G(N)$ approximations to this appendix.

\begin{table}[!ht]
$\begin{array}{|c|r|c|r|r|}
\hline
\mc{3}{|c|}{} & \mc{2}{c|}{\text{Error}} \\ 
\hline
N & \mc{1}{c|}{\text{Exact}} & M & \mc{1}{c|}{\text{HYP } E} & \mc{1}{c|}{\text{HYP } G} \\
\hline
1 &  0.53018 &  3 & -1.68851\x 10^{-04} & -4.61178\x 10^{-05} \\
2 &  2.43002 &  6 &  4.31969\x 10^{-07} &  1.35444\x 10^{-07} \\
3 &  6.15787 &  7 & -1.40070\x 10^{-09} & -4.73745\x 10^{-10} \\
4 & 11.98024 & 10 & -1.09301\x 10^{-11} & -2.45748\x 10^{-12} \\
5 & 20.11115 & 12 &  3.15656\x 10^{-13} &  1.07012\x 10^{-13} \\
6 & 30.73034 & 14 & -5.18497\x 10^{-15} & -1.89058\x 10^{-15} \\
7 & 43.99457 & 16 &  7.31437\x 10^{-17} &  2.76835\x 10^{-17} \\
8 & 60.04387 & 20 & -8.89917\x 10^{-19} & -3.00443\x 10^{-19} \\
9 & 79.00537 & 16 &  6.14869\x 10^{-21} &  1.47877\x 10^{-20} \\
\hline
\end{array}$
\caption{The hyperasymptotic $E(N)$ approximation of Table \ref{tab:SWKB.SC} compared to the hyperasymptotic $G(N)$ approximation of Table \ref{tab:CompGandE} turned into an $E(N)$ approximation by subtracting away half of our best estimate for $\eps_N$: $\eps\CWKB(N)$ in Eq. (\ref{DefCWKB}).  We give the hyperasymptotic $G(N)$ approximation in Eq. (\ref{hyperG}) and use empirical optimal truncation to get the order $M$.}
\label{tab:CompGandE}
\end{table}

For $N = 0$, $2G(0)$ is an estimate of the ground state eigenvalue.  In Table \ref{tab:GSApproxs} we test various approximations to the ground state against the various $G(N)$ approximations developed here.

\begin{table}[!ht]
$\begin{array}{|c|c|r|}
\hline
\text{Approx.} & \text{Order} & \mc{1}{c|}{\text{Error}} \\
\hline
\text{WKB}     & 1 & -0.0352704 \\
\text{SWKB}    & 3 & -0.0037296 \\
\text{G SWKB}  & 1 &  0.0323487 \\
\text{CWKB}    & 1 &  0.0145719 \\
\text{CSWKB}   & 3 &  0.0003043 \\
\text{G CSWKB} & 4 & -0.0024843 \\
\text{E HYP}   & 3 & -0.0001689 \\
\text{G HYP}   & 2 & -0.0040283 \\
\hline
\end{array}$
\caption{Errors in various approximations to $\eps_0 = 0.53018$.  We take the G versions of the SWKB, CSWKB, and HYP approximations at the order of empirical optimal truncation.  The other approximations come from Tables \ref{tab:SCError} and \ref{tab:SWKB.SC}.}
\label{tab:GSApproxs}
\end{table}

%% file: TableGApprox.tex
\begin{table*}[!ht]
$\begin{array}{|c|r|c|c|c|r|r|r|}
\hline
\mc{2}{|c|}{} & \mc{3}{c|}{\text{Order}} & \mc{3}{c|}{\text{Error}}\\
\hline
N & \mc{1}{c|}{\text{Exact}} & \text{SWKB} & \text{CSWKB} & \text{HYP} & \mc{1}{c|}{\text{SWKB}} & \mc{1}{c|}{\text{CSWKB}} & \mc{1}{c|}{\text{HYP}} \\
\hline
 0 &   0.26509 &  3 &  4 &  2 &  3.71006\x 10^{-02} & -1.24217\x 10^{-03} & -2.01413\x 10^{-03} \\
 1 &   1.48010 &  5 &  4 &  3 & -2.03890\x 10^{-03} &  5.66343\x 10^{-05} & -1.97554\x 10^{-05} \\
 2 &   4.29394 &  7 &  6 &  6 &  1.04352\x 10^{-04} & -8.70496\x 10^{-07} &  3.55482\x 10^{-08} \\
 3 &   9.06905 &  9 &  8 &  8 & -5.12434\x 10^{-06} &  1.05384\x 10^{-08} & -4.83320\x 10^{-11} \\
 4 &  16.04570 & 11 & 12 & 10 &  2.43535\x 10^{-07} &  1.17658\x 10^{-10} & -4.52413\x 10^{-12} \\
 5 &  25.42075 & 13 & 14 & 12 & -1.13409\x 10^{-08} & -1.33613\x 10^{-12} &  9.01092\x 10^{-14} \\
 6 &  37.36246 & 15 & 16 & 14 &  5.21047\x 10^{-10} &  1.53515\x 10^{-14} & -1.35614\x 10^{-15} \\
 7 &  52.01922 & 17 & 18 & 16 & -2.37140\x 10^{-11} & -1.78113\x 10^{-16} &  1.83652\x 10^{-17} \\
 8 &  69.52462 & 21 & 20 & 20 &  1.07180\x 10^{-12} &  2.08591\x 10^{-18} & -1.62795\x 10^{-19} \\
 9 &  90.00066 & 23 & 22 & 22 & -4.81857\x 10^{-14} & -2.47297\x 10^{-20} &  8.23128\x 10^{-22} \\
10 & 113.56002 & 25 & 26 & 24 &  2.15726\x 10^{-15} & -2.86261\x 10^{-22} &  6.70277\x 10^{-24} \\
\hline
\end{array}$
\caption{The exact $G(N)$ and the errors of various approximations.  We use least odd order optimal truncation for SWKB and empirical optimal truncation for the other approximations and report the orders at which we truncate.}
\label{tab:GApprox}
\end{table*}

%% file: ConstantsTab.tex
\begin{table*}[!ht]
\scalebox{0.8}{
$\begin{array}{|c|r|r|r|r|r|r|}
\hline
n & \mc{1}{c|}{a_n} & \mc{1}{c|}{b_n} & \mc{1}{c|}{d_n} & \mc{1}{c|}{h_n} & \mc{1}{c|}{h_{n + 20}} & \mc{1}{c|}{c_n} \\
\hline
 1 &-8.567748394584\x10^{-02} &  3.536776513153\x10^{-02} &  5.305164769730\x10^{-02} & -1.666666666667\x10^{-1} &  5.194304272632\x10^{04} &  1.179447262911\x 10^{-01} \\
 2 & 2.148437500000\x10^{-02} & -3.527579746032\x10^{-03} & -1.407238661699\x10^{-03} & -3.794993676877\x10^{-3} &  1.622194035444\x10^{05} &  6.390774747964\x 10^{-03} \\
 3 & 5.239936746010\x10^{-02} & -1.765731366249\x10^{-03} & -2.926108161501\x10^{-03} &  6.596679464131\x10^{-3} & -5.982812586891\x10^{04} &  1.036757028064\x 10^{-05} \\
 4 &-3.188528333391\x10^{-01} &  4.173466791769\x10^{-03} &  4.726357002028\x10^{-04} &  6.104871777506\x10^{-3} & -4.628688644053\x10^{06} & -1.972265116005\x 10^{-03}\\
 5 &-2.906253581665\x10^{+00} &  1.016471233994\x10^{-02} &  1.628857392426\x10^{-02} &  8.012575730396\x10^{-3} & -3.434326216453\x10^{07} & -4.866940504878\x 10^{-05}\\
 6 & 4.149581824006\x10^{+01} & -5.136197691366\x10^{-02} & -4.352082661762\x10^{-03} &  4.764793921301\x10^{-3} & -1.291129925755\x10^{08} &  1.509652803690\x 10^{-02}\\
 7 & 9.106051111453\x10^{+02} & -3.151839648774\x10^{-01} & -4.935113066975\x10^{-01} & -2.526532163520\x10^{-3} &  4.682761458754\x10^{07} & -6.970540753456\x 10^{-05}\\
 8 &-2.759178233251\x10^{+04} &  3.221164312655\x10^{+00} &  1.837863615021\x10^{-01} & -2.047834105176\x10^{-2} &  5.015568432482\x10^{09} & -6.683710532285\x 10^{-01}\\
 9 &-1.086497889469\x10^{+06} &  3.653193762830\x10^{+01} &  5.661119950435\x10^{+01} & -4.760940814614\x10^{-2} &  4.359779213987\x10^{10} & -1.864942552345\x 10^{-04}\\
10 & 5.444198816876\x10^{+07} & -6.046509882828\x10^{+02} & -2.705683388100\x10^{+01} & -5.781448172999\x10^{-2} &  1.916566221097\x10^{11} &  9.795335354117\x 10^{+01}\\
11 & 3.396513559809\x10^{+09} & -1.103195365766\x10^{+04} & -1.698417456463\x10^{+04} &  2.484442856411\x10^{-2} & -6.868413739548\x10^{10} & -1.398393112909\x 10^{+00}\\
12 &-2.575567562301\x10^{+11} &  2.727413882117\x10^{+05} &  9.916174888450\x10^{+03} &  4.073552890690\x10^{-1} & -9.740013068265\x10^{12} & -3.613621803518\x 10^{+04}\\
13 &-2.330666020371\x10^{+13} &  7.291976034200\x10^{+06} &  1.117814650219\x10^{+07} &  1.481124311328\x10^{+0} & -9.705364249021\x10^{13} &  7.386689584732\x 10^{+02}\\
14 & 2.482592926162\x10^{+15} & -2.510854389322\x10^{+08} & -7.711376112560\x10^{+06} &  2.741314325143\x10^{+0} & -4.884300202923\x10^{14} &  2.816398059630\x 10^{+07}\\
15 & 3.075586665318\x10^{+17} & -9.255480436013\x10^{+09} & -1.414428669791\x10^{+10} & -1.081551161491\x10^{+0} &  1.734725923736\x10^{14} & -8.442791957114\x 10^{+05}\\
16 &-4.384018949045\x10^{+19} &  4.238678892387\x10^{+11} &  1.125769296301\x10^{+10} & -3.528471843003\x10^{+1} &  3.145775486985\x10^{16} & -4.121307364131\x 10^{+10}\\
17 &-7.123931840874\x10^{+21} &  2.060168174148\x10^{+13} &  3.141070779556\x10^{+13} & -1.719162719101\x10^{+2} &  3.534910367892\x10^{17} &  1.580982245534\x 10^{+09}\\
18 & 1.309156462631\x10^{+24} & -1.210778641220\x10^{+15} & -2.833222450361\x10^{+13} & -4.274928378220\x10^{+2} &  2.003996635297\x10^{18} &  1.038956384366\x 10^{+14}\\
19 & 2.701473930086\x10^{+26} & -7.503088536990\x10^{+16} & -1.141906853721\x10^{+17} &  1.616892923721\x10^{+2} & -7.067807271510\x10^{17} & -4.901331150371\x 10^{+12}\\
20 &-6.219943513829\x10^{+28} &  5.505039532830\x10^{+18} &  1.151129625648\x10^{+17} &  8.443442098119\x10^{+3} & -1.595548697681\x10^{20} & -4.227171913689\x 10^{+17}\\
\hline
\end{array}$}
\caption{We present to 13 digits the first 21 examples from each set of coefficients used in this paper (remember that the zeroth order coefficient for each set is 1).  We give more coefficients and more digits in the supplemental info (Tables S2, S5, S8, S11, and S14).}
\label{tab:EigCoeff}
\end{table*}

%% file: arXivRev.bbl
\begin{thebibliography}{60}%
	\makeatletter
	\providecommand \@ifxundefined [1]{%
		\@ifx{#1\undefined}
	}%
	\providecommand \@ifnum [1]{%
		\ifnum #1\expandafter \@firstoftwo
		\else \expandafter \@secondoftwo
		\fi
	}%
	\providecommand \@ifx [1]{%
		\ifx #1\expandafter \@firstoftwo
		\else \expandafter \@secondoftwo
		\fi
	}%
	\providecommand \natexlab [1]{#1}%
	\providecommand \enquote  [1]{``#1''}%
	\providecommand \bibnamefont  [1]{#1}%
	\providecommand \bibfnamefont [1]{#1}%
	\providecommand \citenamefont [1]{#1}%
	\providecommand \href@noop [0]{\@secondoftwo}%
	\providecommand \href [0]{\begingroup \@sanitize@url \@href}%
	\providecommand \@href[1]{\@@startlink{#1}\@@href}%
	\providecommand \@@href[1]{\endgroup#1\@@endlink}%
	\providecommand \@sanitize@url [0]{\catcode `\\12\catcode `\$12\catcode
		`\&12\catcode `\#12\catcode `\^12\catcode `\_12\catcode `\%12\relax}%
	\providecommand \@@startlink[1]{}%
	\providecommand \@@endlink[0]{}%
	\providecommand \url  [0]{\begingroup\@sanitize@url \@url }%
	\providecommand \@url [1]{\endgroup\@href {#1}{\urlprefix }}%
	\providecommand \urlprefix  [0]{URL }%
	\providecommand \Eprint [0]{\href }%
	\providecommand \doibase [0]{https://doi.org/}%
	\providecommand \selectlanguage [0]{\@gobble}%
	\providecommand \bibinfo  [0]{\@secondoftwo}%
	\providecommand \bibfield  [0]{\@secondoftwo}%
	\providecommand \translation [1]{[#1]}%
	\providecommand \BibitemOpen [0]{}%
	\providecommand \bibitemStop [0]{}%
	\providecommand \bibitemNoStop [0]{.\EOS\space}%
	\providecommand \EOS [0]{\spacefactor3000\relax}%
	\providecommand \BibitemShut  [1]{\csname bibitem#1\endcsname}%
	\let\auto@bib@innerbib\@empty
	\bibitem [{\citenamefont {Hartree}\ and\ \citenamefont {Hartree}(1935)}]{HH35}%
	\BibitemOpen
	\bibfield  {author} {\bibinfo {author} {\bibfnamefont {D.~R.}\ \bibnamefont
			{Hartree}}\ and\ \bibinfo {author} {\bibfnamefont {W.}~\bibnamefont
			{Hartree}},\ }\bibfield  {title} {\bibinfo {title} {Self-consistent field,
			with exchange, for beryllium},\ }\href
	{https://doi.org/10.1098/rspa.1935.0085} {\bibfield  {journal} {\bibinfo
			{journal} {Proceedings of the Royal Society of London. Series A -
				Mathematical and Physical Sciences}\ }\textbf {\bibinfo {volume} {150}},\
		\bibinfo {pages} {9} (\bibinfo {year} {1935})}\BibitemShut {NoStop}%
	\bibitem [{\citenamefont {Szabo}\ and\ \citenamefont {Ostlund}(1996)}]{SO96}%
	\BibitemOpen
	\bibfield  {author} {\bibinfo {author} {\bibfnamefont {A.}~\bibnamefont
			{Szabo}}\ and\ \bibinfo {author} {\bibfnamefont {N.~S.}\ \bibnamefont
			{Ostlund}},\ }\href@noop {} {\emph {\bibinfo {title} {Modern quantum
				chemistry}}}\ (\bibinfo  {publisher} {Dover Publications},\ \bibinfo {year}
	{1996})\BibitemShut {NoStop}%
	\bibitem [{\citenamefont {March}\ and\ \citenamefont {Plaskett}(1956)}]{MP56}%
	\BibitemOpen
	\bibfield  {author} {\bibinfo {author} {\bibfnamefont {N.~H.}\ \bibnamefont
			{March}}\ and\ \bibinfo {author} {\bibfnamefont {J.~S.}\ \bibnamefont
			{Plaskett}},\ }\bibfield  {title} {\bibinfo {title} {The relation between the
			wentzel-kramers-brillouin and the thomas-fermi approximations},\ }\href
	{https://doi.org/10.1098/rspa.1956.0094} {\bibfield  {journal} {\bibinfo
			{journal} {Proceedings of the Royal Society of London. Series A. Mathematical
				and Physical Sciences}\ }\textbf {\bibinfo {volume} {235}},\ \bibinfo {pages}
		{419} (\bibinfo {year} {1956})}\BibitemShut {NoStop}%
	\bibitem [{\citenamefont {Englert}(1988)}]{E88}%
	\BibitemOpen
	\bibfield  {author} {\bibinfo {author} {\bibfnamefont {B.-G.}\ \bibnamefont
			{Englert}},\ }\href@noop {} {\emph {\bibinfo {title} {Semiclassical Theory of
				Atoms}}}\ (\bibinfo  {publisher} {Springer},\ \bibinfo {address} {Verlag
		Berlin Heidelberg},\ \bibinfo {year} {1988})\BibitemShut {NoStop}%
	\bibitem [{\citenamefont {Perdew}\ \emph {et~al.}(2006)\citenamefont {Perdew},
		\citenamefont {Constantin}, \citenamefont {Sagvolden},\ and\ \citenamefont
		{Burke}}]{PCSB06}%
	\BibitemOpen
	\bibfield  {author} {\bibinfo {author} {\bibfnamefont {J.~P.}\ \bibnamefont
			{Perdew}}, \bibinfo {author} {\bibfnamefont {L.~A.}\ \bibnamefont
			{Constantin}}, \bibinfo {author} {\bibfnamefont {E.}~\bibnamefont
			{Sagvolden}},\ and\ \bibinfo {author} {\bibfnamefont {K.}~\bibnamefont
			{Burke}},\ }\bibfield  {title} {\bibinfo {title} {Relevance of the slowly
			varying electron gas to atoms, molecules, and solids},\ }\href
	{https://doi.org/10.1103/PhysRevLett.97.223002} {\bibfield  {journal}
		{\bibinfo  {journal} {Phys. Rev. Lett.}\ }\textbf {\bibinfo {volume} {97}},\
		\bibinfo {pages} {223002} (\bibinfo {year} {2006})}\BibitemShut {NoStop}%
	\bibitem [{\citenamefont {Elliott}\ and\ \citenamefont {Burke}(2009)}]{EB09}%
	\BibitemOpen
	\bibfield  {author} {\bibinfo {author} {\bibfnamefont {P.}~\bibnamefont
			{Elliott}}\ and\ \bibinfo {author} {\bibfnamefont {K.}~\bibnamefont
			{Burke}},\ }\bibfield  {title} {\bibinfo {title} {Non-empirical derivation of
			the parameter in the b88 exchange functional},\ }\href
	{https://doi.org/10.1139/V09-095} {\bibfield  {journal} {\bibinfo  {journal}
			{Can. J. Chem. Ecol.}\ }\textbf {\bibinfo {volume} {87}},\ \bibinfo {pages}
		{1485} (\bibinfo {year} {2009})}\BibitemShut {NoStop}%
	\bibitem [{\citenamefont {Burke}\ \emph {et~al.}(2016)\citenamefont {Burke},
		\citenamefont {Cancio}, \citenamefont {Gould},\ and\ \citenamefont
		{Pittalis}}]{BCGP16}%
	\BibitemOpen
	\bibfield  {author} {\bibinfo {author} {\bibfnamefont {K.}~\bibnamefont
			{Burke}}, \bibinfo {author} {\bibfnamefont {A.}~\bibnamefont {Cancio}},
		\bibinfo {author} {\bibfnamefont {T.}~\bibnamefont {Gould}},\ and\ \bibinfo
		{author} {\bibfnamefont {S.}~\bibnamefont {Pittalis}},\ }\bibfield  {title}
	{\bibinfo {title} {Locality of correlation in density functional theory},\
	}\href {https://doi.org/10.1063/1.4959126} {\bibfield  {journal} {\bibinfo
			{journal} {The Journal of Chemical Physics}\ }\textbf {\bibinfo {volume}
			{145}},\ \bibinfo {pages} {054112} (\bibinfo {year} {2016})}\BibitemShut
	{NoStop}%
	\bibitem [{\citenamefont {Cancio}\ \emph {et~al.}(2018)\citenamefont {Cancio},
		\citenamefont {Chen}, \citenamefont {Krull},\ and\ \citenamefont
		{Burke}}]{CCKB18}%
	\BibitemOpen
	\bibfield  {author} {\bibinfo {author} {\bibfnamefont {A.}~\bibnamefont
			{Cancio}}, \bibinfo {author} {\bibfnamefont {G.~P.}\ \bibnamefont {Chen}},
		\bibinfo {author} {\bibfnamefont {B.~T.}\ \bibnamefont {Krull}},\ and\
		\bibinfo {author} {\bibfnamefont {K.}~\bibnamefont {Burke}},\ }\bibfield
	{title} {\bibinfo {title} {Fitting a round peg into a round hole:
			Asymptotically correcting the generalized gradient approximation for
			correlation},\ }\href {https://doi.org/10.1063/1.5021597} {\bibfield
		{journal} {\bibinfo  {journal} {The Journal of Chemical Physics}\ }\textbf
		{\bibinfo {volume} {149}},\ \bibinfo {pages} {084116} (\bibinfo {year}
		{2018})}\BibitemShut {NoStop}%
	\bibitem [{\citenamefont {Sun}\ \emph {et~al.}(2015)\citenamefont {Sun},
		\citenamefont {Ruzsinszky},\ and\ \citenamefont {Perdew}}]{SRP15}%
	\BibitemOpen
	\bibfield  {author} {\bibinfo {author} {\bibfnamefont {J.}~\bibnamefont
			{Sun}}, \bibinfo {author} {\bibfnamefont {A.}~\bibnamefont {Ruzsinszky}},\
		and\ \bibinfo {author} {\bibfnamefont {J.~P.}\ \bibnamefont {Perdew}},\
	}\bibfield  {title} {\bibinfo {title} {Strongly constrained and appropriately
			normed semilocal density functional},\ }\href
	{https://doi.org/10.1103/PhysRevLett.115.036402} {\bibfield  {journal}
		{\bibinfo  {journal} {Phys. Rev. Lett.}\ }\textbf {\bibinfo {volume} {115}},\
		\bibinfo {pages} {036402} (\bibinfo {year} {2015})}\BibitemShut {NoStop}%
	\bibitem [{\citenamefont {Perdew}\ \emph {et~al.}(2008)\citenamefont {Perdew},
		\citenamefont {Ruzsinszky}, \citenamefont {Csonka}, \citenamefont {Vydrov},
		\citenamefont {Scuseria}, \citenamefont {Constantin}, \citenamefont {Zhou},\
		and\ \citenamefont {Burke}}]{PRCV08}%
	\BibitemOpen
	\bibfield  {author} {\bibinfo {author} {\bibfnamefont {J.~P.}\ \bibnamefont
			{Perdew}}, \bibinfo {author} {\bibfnamefont {A.}~\bibnamefont {Ruzsinszky}},
		\bibinfo {author} {\bibfnamefont {G.~I.}\ \bibnamefont {Csonka}}, \bibinfo
		{author} {\bibfnamefont {O.~A.}\ \bibnamefont {Vydrov}}, \bibinfo {author}
		{\bibfnamefont {G.~E.}\ \bibnamefont {Scuseria}}, \bibinfo {author}
		{\bibfnamefont {L.~A.}\ \bibnamefont {Constantin}}, \bibinfo {author}
		{\bibfnamefont {X.}~\bibnamefont {Zhou}},\ and\ \bibinfo {author}
		{\bibfnamefont {K.}~\bibnamefont {Burke}},\ }\bibfield  {title} {\bibinfo
		{title} {Restoring the density-gradient expansion for exchange in solids and
			surfaces},\ }\href {https://doi.org/10.1103/PhysRevLett.100.136406}
	{\bibfield  {journal} {\bibinfo  {journal} {Phys. Rev. Lett.}\ }\textbf
		{\bibinfo {volume} {100}},\ \bibinfo {pages} {136406} (\bibinfo {year}
		{2008})}\BibitemShut {NoStop}%
	\bibitem [{\citenamefont {Burke}(2020{\natexlab{a}})}]{B20}%
	\BibitemOpen
	\bibfield  {author} {\bibinfo {author} {\bibfnamefont {K.}~\bibnamefont
			{Burke}},\ }\bibfield  {title} {\bibinfo {title} {Leading correction to the
			local density approximation of the kinetic energy in one dimension},\ }\href
	{https://doi.org/10.1063/5.0002287} {\bibfield  {journal} {\bibinfo
			{journal} {The Journal of Chemical Physics}\ }\textbf {\bibinfo {volume}
			{152}},\ \bibinfo {pages} {081102} (\bibinfo {year}
		{2020}{\natexlab{a}})}\BibitemShut {NoStop}%
	\bibitem [{\citenamefont {Burke}(2020{\natexlab{b}})}]{B20b}%
	\BibitemOpen
	\bibfield  {author} {\bibinfo {author} {\bibfnamefont {K.}~\bibnamefont
			{Burke}},\ }\bibfield  {title} {\bibinfo {title} {Deriving approximate
			density functionals with asymptotics},\ }\href
	{https://doi.org/10.1039/D0FD00057D} {\bibfield  {journal} {\bibinfo
			{journal} {Faraday Discuss.}\ }\textbf {\bibinfo {volume} {224}},\ \bibinfo
		{pages} {98} (\bibinfo {year} {2020}{\natexlab{b}})}\BibitemShut {NoStop}%
	\bibitem [{\citenamefont {Berry}\ and\ \citenamefont {Burke}(2020)}]{BB20}%
	\BibitemOpen
	\bibfield  {author} {\bibinfo {author} {\bibfnamefont {M.~V.}\ \bibnamefont
			{Berry}}\ and\ \bibinfo {author} {\bibfnamefont {K.}~\bibnamefont {Burke}},\
	}\bibfield  {title} {\bibinfo {title} {Exact and approximate energy sums in
			potential wells},\ }\href {https://doi.org/10.1088/1751-8121/ab69a6}
	{\bibfield  {journal} {\bibinfo  {journal} {Journal of Physics A:
				Mathematical and Theoretical}\ }\textbf {\bibinfo {volume} {53}},\ \bibinfo
		{pages} {095203} (\bibinfo {year} {2020})}\BibitemShut {NoStop}%
	\bibitem [{\citenamefont {Boyd}(1999)}]{B99}%
	\BibitemOpen
	\bibfield  {author} {\bibinfo {author} {\bibfnamefont {J.~P.}\ \bibnamefont
			{Boyd}},\ }\bibfield  {title} {\bibinfo {title} {The devil's invention:
			Asymptotic, superasymptotic and hyperasymptotic series},\ }\href
	{https://doi.org/10.1023/A:1006145903624} {\bibfield  {journal} {\bibinfo
			{journal} {Acta Applicandae Mathematica}\ }\textbf {\bibinfo {volume} {56}},\
		\bibinfo {pages} {1} (\bibinfo {year} {1999})}\BibitemShut {NoStop}%
	\bibitem [{\citenamefont {Berry}(1991)}]{B91}%
	\BibitemOpen
	\bibfield  {author} {\bibinfo {author} {\bibfnamefont {M.}~\bibnamefont
			{Berry}},\ }\bibfield  {title} {\bibinfo {title} {Asymptotics,
			superasymptotics, hyperasymptotics...},\ }in\ \href
	{https://doi.org/10.1007/978-1-4757-0435-8_1} {\emph {\bibinfo {booktitle}
			{Asymptotics beyond All Orders}}},\ \bibinfo {editor} {edited by\ \bibinfo
		{editor} {\bibfnamefont {H.}~\bibnamefont {Segur}}, \bibinfo {editor}
		{\bibfnamefont {S.}~\bibnamefont {Tanveer}},\ and\ \bibinfo {editor}
		{\bibfnamefont {H.}~\bibnamefont {Levine}}}\ (\bibinfo  {publisher}
	{Springer},\ \bibinfo {address} {Boston},\ \bibinfo {year} {1991})\ pp.\
	\bibinfo {pages} {1--14}\BibitemShut {NoStop}%
	\bibitem [{\citenamefont {HELLER}(2018)}]{H18}%
	\BibitemOpen
	\bibfield  {author} {\bibinfo {author} {\bibfnamefont {E.~J.}\ \bibnamefont
			{HELLER}},\ }\href {http://www.jstor.org/stable/j.ctvc77gwd} {\emph {\bibinfo
			{title} {The Semiclassical Way to Dynamics and Spectroscopy}}}\ (\bibinfo
	{publisher} {Princeton University Press},\ \bibinfo {year}
	{2018})\BibitemShut {NoStop}%
	\bibitem [{\citenamefont {Child}(2014)}]{C14}%
	\BibitemOpen
	\bibfield  {author} {\bibinfo {author} {\bibfnamefont {M.~S.}\ \bibnamefont
			{Child}},\ }\href {https://doi.org/10.1093/acprof:oso/9780199672981.001.0001}
	{\emph {\bibinfo {title} {Semiclassical Mechanics with Molecular
				Applications}}},\ \bibinfo {edition} {2nd}\ ed.\ (\bibinfo  {publisher}
	{Oxford University Press},\ \bibinfo {address} {New York},\ \bibinfo {year}
	{2014})\BibitemShut {NoStop}%
	\bibitem [{\citenamefont {Okun}\ and\ \citenamefont
		{Burke}(2021{\natexlab{a}})}]{OB21}%
	\BibitemOpen
	\bibfield  {author} {\bibinfo {author} {\bibfnamefont {P.}~\bibnamefont
			{Okun}}\ and\ \bibinfo {author} {\bibfnamefont {K.}~\bibnamefont {Burke}},\
	}\bibfield  {title} {\bibinfo {title} {Semiclassics: The hidden theory behind
			the success of dft}} (\bibinfo {year} {2021}{\natexlab{a}}),\ \bibinfo {note}
	{arXiv:2105.04384}\BibitemShut {NoStop}%
	\bibitem [{\citenamefont {Cangi}\ \emph {et~al.}(2011)\citenamefont {Cangi},
		\citenamefont {Lee}, \citenamefont {Elliott}, \citenamefont {Burke},\ and\
		\citenamefont {Gross}}]{CLEB11}%
	\BibitemOpen
	\bibfield  {author} {\bibinfo {author} {\bibfnamefont {A.}~\bibnamefont
			{Cangi}}, \bibinfo {author} {\bibfnamefont {D.}~\bibnamefont {Lee}}, \bibinfo
		{author} {\bibfnamefont {P.}~\bibnamefont {Elliott}}, \bibinfo {author}
		{\bibfnamefont {K.}~\bibnamefont {Burke}},\ and\ \bibinfo {author}
		{\bibfnamefont {E.~K.~U.}\ \bibnamefont {Gross}},\ }\bibfield  {title}
	{\bibinfo {title} {Electronic structure via potential functional
			approximations},\ }\href {https://doi.org/10.1103/PhysRevLett.106.236404}
	{\bibfield  {journal} {\bibinfo  {journal} {Phys. Rev. Lett.}\ }\textbf
		{\bibinfo {volume} {106}},\ \bibinfo {pages} {236404} (\bibinfo {year}
		{2011})}\BibitemShut {NoStop}%
	\bibitem [{\citenamefont {Cangi}\ \emph {et~al.}(2013)\citenamefont {Cangi},
		\citenamefont {Gross},\ and\ \citenamefont {Burke}}]{CGB13}%
	\BibitemOpen
	\bibfield  {author} {\bibinfo {author} {\bibfnamefont {A.}~\bibnamefont
			{Cangi}}, \bibinfo {author} {\bibfnamefont {E.~K.~U.}\ \bibnamefont
			{Gross}},\ and\ \bibinfo {author} {\bibfnamefont {K.}~\bibnamefont {Burke}},\
	}\bibfield  {title} {\bibinfo {title} {Potential functionals versus density
			functionals},\ }\href {https://doi.org/10.1103/PhysRevA.88.062505} {\bibfield
		{journal} {\bibinfo  {journal} {Phys. Rev. A}\ }\textbf {\bibinfo {volume}
			{88}},\ \bibinfo {pages} {062505} (\bibinfo {year} {2013})}\BibitemShut
	{NoStop}%
	\bibitem [{\citenamefont {Lign{\`e}res}\ and\ \citenamefont
		{Carter}(2005)}]{LC05}%
	\BibitemOpen
	\bibfield  {author} {\bibinfo {author} {\bibfnamefont {V.~L.}\ \bibnamefont
			{Lign{\`e}res}}\ and\ \bibinfo {author} {\bibfnamefont {E.~A.}\ \bibnamefont
			{Carter}},\ }\bibinfo {title} {An introduction to orbital-free density
		functional theory},\ in\ \href {https://doi.org/10.1007/978-1-4020-3286-8_9}
	{\emph {\bibinfo {booktitle} {Handbook of Materials Modeling: Methods}}},\
	\bibinfo {editor} {edited by\ \bibinfo {editor} {\bibfnamefont
			{S.}~\bibnamefont {Yip}}}\ (\bibinfo  {publisher} {Springer Netherlands},\
	\bibinfo {address} {Dordrecht},\ \bibinfo {year} {2005})\ pp.\ \bibinfo
	{pages} {137--148}\BibitemShut {NoStop}%
	\bibitem [{\citenamefont {Bender}\ and\ \citenamefont {Orszag}(1999)}]{BO99}%
	\BibitemOpen
	\bibfield  {author} {\bibinfo {author} {\bibfnamefont {C.~M.}\ \bibnamefont
			{Bender}}\ and\ \bibinfo {author} {\bibfnamefont {S.~A.}\ \bibnamefont
			{Orszag}},\ }\href {https://doi.org/10.1007/978-1-4757-3069-2} {\emph
		{\bibinfo {title} {Advanced Mathematical Methods for Scientists and Engineers
				I: Asymptotic Methods and Perturbation Theory}}}\ (\bibinfo  {publisher}
	{Springer},\ \bibinfo {address} {Verlag New York},\ \bibinfo {year}
	{1999})\BibitemShut {NoStop}%
	\bibitem [{\citenamefont {Bender}\ and\ \citenamefont {Jones}(2012)}]{BJ12}%
	\BibitemOpen
	\bibfield  {author} {\bibinfo {author} {\bibfnamefont {C.~M.}\ \bibnamefont
			{Bender}}\ and\ \bibinfo {author} {\bibfnamefont {H.~F.}\ \bibnamefont
			{Jones}},\ }\bibfield  {title} {\bibinfo {title} {{WKB} analysis of
			pt-symmetric sturm{\textendash}liouville problems},\ }\href
	{https://doi.org/10.1088/1751-8113/45/44/444004} {\bibfield  {journal}
		{\bibinfo  {journal} {Journal of Physics A: Mathematical and Theoretical}\
		}\textbf {\bibinfo {volume} {45}},\ \bibinfo {pages} {444004} (\bibinfo
		{year} {2012})}\BibitemShut {NoStop}%
	\bibitem [{\citenamefont {Dutt}\ \emph {et~al.}(1993)\citenamefont {Dutt},
		\citenamefont {Gangopadhyaya}, \citenamefont {Khare}, \citenamefont
		{Pagnamenta},\ and\ \citenamefont {Sukhatme}}]{DGKS93}%
	\BibitemOpen
	\bibfield  {author} {\bibinfo {author} {\bibfnamefont {R.}~\bibnamefont
			{Dutt}}, \bibinfo {author} {\bibfnamefont {A.}~\bibnamefont {Gangopadhyaya}},
		\bibinfo {author} {\bibfnamefont {A.}~\bibnamefont {Khare}}, \bibinfo
		{author} {\bibfnamefont {A.}~\bibnamefont {Pagnamenta}},\ and\ \bibinfo
		{author} {\bibfnamefont {U.}~\bibnamefont {Sukhatme}},\ }\bibfield  {title}
	{\bibinfo {title} {Semiclassical approach to quantum-mechanical problems with
			broken supersymmetry},\ }\href {https://doi.org/10.1103/PhysRevA.48.1845}
	{\bibfield  {journal} {\bibinfo  {journal} {Phys. Rev. A}\ }\textbf {\bibinfo
			{volume} {48}},\ \bibinfo {pages} {1845} (\bibinfo {year}
		{1993})}\BibitemShut {NoStop}%
	\bibitem [{\citenamefont {Cooper}\ \emph {et~al.}(1995)\citenamefont {Cooper},
		\citenamefont {Khare},\ and\ \citenamefont {Sukhatme}}]{CKS95}%
	\BibitemOpen
	\bibfield  {author} {\bibinfo {author} {\bibfnamefont {F.}~\bibnamefont
			{Cooper}}, \bibinfo {author} {\bibfnamefont {A.}~\bibnamefont {Khare}},\ and\
		\bibinfo {author} {\bibfnamefont {U.}~\bibnamefont {Sukhatme}},\ }\bibfield
	{title} {\bibinfo {title} {Supersymmetry and quantum mechanics},\ }\href
	{https://doi.org/https://doi.org/10.1016/0370-1573(94)00080-M} {\bibfield
		{journal} {\bibinfo  {journal} {Physics Reports}\ }\textbf {\bibinfo {volume}
			{251}},\ \bibinfo {pages} {267 } (\bibinfo {year} {1995})}\BibitemShut
	{NoStop}%
	\bibitem [{\citenamefont {Aniceto}\ \emph {et~al.}(2019)\citenamefont
		{Aniceto}, \citenamefont {Basar},\ and\ \citenamefont {Schiappa}}]{ABS19}%
	\BibitemOpen
	\bibfield  {author} {\bibinfo {author} {\bibfnamefont {I.}~\bibnamefont
			{Aniceto}}, \bibinfo {author} {\bibfnamefont {G.}~\bibnamefont {Basar}},\
		and\ \bibinfo {author} {\bibfnamefont {R.}~\bibnamefont {Schiappa}},\
	}\bibfield  {title} {\bibinfo {title} {A primer on resurgent transseries and
			their asymptotics},\ }\href
	{https://doi.org/https://doi.org/10.1016/j.physrep.2019.02.003} {\bibfield
		{journal} {\bibinfo  {journal} {Physics Reports}\ }\textbf {\bibinfo {volume}
			{809}},\ \bibinfo {pages} {1} (\bibinfo {year} {2019})}\BibitemShut {NoStop}%
	\bibitem [{{\relax DLMF}()}]{DLMF}%
	\BibitemOpen
	{\relax DLMF},\ \href {http://dlmf.nist.gov/} {\bibinfo {title} {{\it NIST
				Digital Library of Mathematical Functions}}},\ \bibinfo {howpublished}
	{http://dlmf.nist.gov/, Release 1.1.3 of 2021-09-15},\ \bibinfo {note}
	{f.~W.~J. Olver, A.~B. {Olde Daalhuis}, D.~W. Lozier, B.~I. Schneider, R.~F.
		Boisvert, C.~W. Clark, B.~R. Miller, B.~V. Saunders, H.~S. Cohl, and M.~A.
		McClain, eds.}\BibitemShut {Stop}%
	\bibitem [{\citenamefont {Lay}(1997)}]{L97}%
	\BibitemOpen
	\bibfield  {author} {\bibinfo {author} {\bibfnamefont {W.}~\bibnamefont
			{Lay}},\ }\bibfield  {title} {\bibinfo {title} {The quartic oscillator},\
	}\href {https://doi.org/10.1063/1.531857} {\bibfield  {journal} {\bibinfo
			{journal} {Journal of Mathematical Physics}\ }\textbf {\bibinfo {volume}
			{38}},\ \bibinfo {pages} {639} (\bibinfo {year} {1997})},\ \Eprint
	{https://arxiv.org/abs/https://doi.org/10.1063/1.531857}
	{https://doi.org/10.1063/1.531857} \BibitemShut {NoStop}%
	\bibitem [{\citenamefont {Bay}\ and\ \citenamefont {Lay}(1997)}]{BL97}%
	\BibitemOpen
	\bibfield  {author} {\bibinfo {author} {\bibfnamefont {K.}~\bibnamefont
			{Bay}}\ and\ \bibinfo {author} {\bibfnamefont {W.}~\bibnamefont {Lay}},\
	}\bibfield  {title} {\bibinfo {title} {The spectrum of the quartic
			oscillator},\ }\href {https://doi.org/10.1063/1.531962} {\bibfield  {journal}
		{\bibinfo  {journal} {Journal of Mathematical Physics}\ }\textbf {\bibinfo
			{volume} {38}},\ \bibinfo {pages} {2127} (\bibinfo {year}
		{1997})}\BibitemShut {NoStop}%
	\bibitem [{\citenamefont {Liverts}\ \emph {et~al.}(2006)\citenamefont
		{Liverts}, \citenamefont {Mandelzweig},\ and\ \citenamefont
		{Tabakin}}]{LMT06}%
	\BibitemOpen
	\bibfield  {author} {\bibinfo {author} {\bibfnamefont {E.~Z.}\ \bibnamefont
			{Liverts}}, \bibinfo {author} {\bibfnamefont {V.~B.}\ \bibnamefont
			{Mandelzweig}},\ and\ \bibinfo {author} {\bibfnamefont {F.}~\bibnamefont
			{Tabakin}},\ }\bibfield  {title} {\bibinfo {title} {Analytic calculation of
			energies and wave functions of the quartic and pure quartic oscillators},\
	}\href {https://doi.org/10.1063/1.2209769} {\bibfield  {journal} {\bibinfo
			{journal} {Journal of Mathematical Physics}\ }\textbf {\bibinfo {volume}
			{47}},\ \bibinfo {pages} {062109} (\bibinfo {year} {2006})}\BibitemShut
	{NoStop}%
	\bibitem [{\citenamefont {Loeffel}\ \emph {et~al.}(1969)\citenamefont
		{Loeffel}, \citenamefont {Martin}, \citenamefont {Simon},\ and\ \citenamefont
		{Wightman}}]{LMSW69}%
	\BibitemOpen
	\bibfield  {author} {\bibinfo {author} {\bibfnamefont {J.}~\bibnamefont
			{Loeffel}}, \bibinfo {author} {\bibfnamefont {A.}~\bibnamefont {Martin}},
		\bibinfo {author} {\bibfnamefont {B.}~\bibnamefont {Simon}},\ and\ \bibinfo
		{author} {\bibfnamefont {A.}~\bibnamefont {Wightman}},\ }\bibfield  {title}
	{\bibinfo {title} {Pade approximants and the anharmonic oscillator},\ }\href
	{https://doi.org/https://doi.org/10.1016/0370-2693(69)90087-2} {\bibfield
		{journal} {\bibinfo  {journal} {Physics Letters B}\ }\textbf {\bibinfo
			{volume} {30}},\ \bibinfo {pages} {656} (\bibinfo {year} {1969})}\BibitemShut
	{NoStop}%
	\bibitem [{\citenamefont {Simon}\ and\ \citenamefont {Dicke}(1970)}]{S69}%
	\BibitemOpen
	\bibfield  {author} {\bibinfo {author} {\bibfnamefont {B.}~\bibnamefont
			{Simon}}\ and\ \bibinfo {author} {\bibfnamefont {A.}~\bibnamefont {Dicke}},\
	}\bibfield  {title} {\bibinfo {title} {Coupling constant analyticity for the
			anharmonic oscillator},\ }\href
	{https://doi.org/https://doi.org/10.1016/0003-4916(70)90240-X} {\bibfield
		{journal} {\bibinfo  {journal} {Annals of Physics}\ }\textbf {\bibinfo
			{volume} {58}},\ \bibinfo {pages} {76} (\bibinfo {year} {1970})}\BibitemShut
	{NoStop}%
	\bibitem [{\citenamefont {Bender}\ and\ \citenamefont {Wu}(1969)}]{BW69}%
	\BibitemOpen
	\bibfield  {author} {\bibinfo {author} {\bibfnamefont {C.~M.}\ \bibnamefont
			{Bender}}\ and\ \bibinfo {author} {\bibfnamefont {T.~T.}\ \bibnamefont
			{Wu}},\ }\bibfield  {title} {\bibinfo {title} {Anharmonic oscillator},\
	}\href {https://doi.org/10.1103/PhysRev.184.1231} {\bibfield  {journal}
		{\bibinfo  {journal} {Phys. Rev.}\ }\textbf {\bibinfo {volume} {184}},\
		\bibinfo {pages} {1231} (\bibinfo {year} {1969})}\BibitemShut {NoStop}%
	\bibitem [{\citenamefont {GRAFFI}\ \emph {et~al.}(1990)\citenamefont {GRAFFI},
		\citenamefont {GRECCHI},\ and\ \citenamefont {SIMON}}]{GGS70}%
	\BibitemOpen
	\bibfield  {author} {\bibinfo {author} {\bibfnamefont {S.}~\bibnamefont
			{GRAFFI}}, \bibinfo {author} {\bibfnamefont {V.}~\bibnamefont {GRECCHI}},\
		and\ \bibinfo {author} {\bibfnamefont {B.}~\bibnamefont {SIMON}},\ }\bibfield
	{title} {\bibinfo {title} {Borel summability: Application to the anharmonic
			oscillator},\ }in\ \href
	{https://doi.org/https://doi.org/10.1016/B978-0-444-88597-5.50033-8} {\emph
		{\bibinfo {booktitle} {Large-Order Behaviour of Perturbation Theory}}},\
	\bibinfo {series} {Current Physics–Sources and Comments}, Vol.~\bibinfo
	{volume} {7},\ \bibinfo {editor} {edited by\ \bibinfo {editor} {\bibfnamefont
			{J.}~\bibnamefont {{LE GUILLOU}}}\ and\ \bibinfo {editor} {\bibfnamefont
			{J.}~\bibnamefont {ZINN-JUSTIN}}}\ (\bibinfo  {publisher} {Elsevier},\
	\bibinfo {year} {1990})\ pp.\ \bibinfo {pages} {240--243}\BibitemShut
	{NoStop}%
	\bibitem [{\citenamefont {Hioe}\ and\ \citenamefont {Montroll}(1975)}]{HM75}%
	\BibitemOpen
	\bibfield  {author} {\bibinfo {author} {\bibfnamefont {F.~T.}\ \bibnamefont
			{Hioe}}\ and\ \bibinfo {author} {\bibfnamefont {E.~W.}\ \bibnamefont
			{Montroll}},\ }\bibfield  {title} {\bibinfo {title} {Quantum theory of
			anharmonic oscillators. i. energy levels of oscillators with positive quartic
			anharmonicity},\ }\href {https://doi.org/10.1063/1.522747} {\bibfield
		{journal} {\bibinfo  {journal} {Journal of Mathematical Physics}\ }\textbf
		{\bibinfo {volume} {16}},\ \bibinfo {pages} {1945} (\bibinfo {year}
		{1975})}\BibitemShut {NoStop}%
	\bibitem [{\citenamefont {Balian}\ \emph {et~al.}(1979)\citenamefont {Balian},
		\citenamefont {Parisi},\ and\ \citenamefont {Voros}}]{BPV79}%
	\BibitemOpen
	\bibfield  {author} {\bibinfo {author} {\bibfnamefont {R.}~\bibnamefont
			{Balian}}, \bibinfo {author} {\bibfnamefont {G.}~\bibnamefont {Parisi}},\
		and\ \bibinfo {author} {\bibfnamefont {A.}~\bibnamefont {Voros}},\ }\bibfield
	{title} {\bibinfo {title} {Quartic oscillator},\ }in\ \href
	{https://doi.org/https://doi.org/10.1007/3-540-09532-2_85} {\emph {\bibinfo
			{booktitle} {Feynman Path Integrals}}},\ \bibinfo {editor} {edited by\
		\bibinfo {editor} {\bibfnamefont {S.}~\bibnamefont {Albeverio}}, \bibinfo
		{editor} {\bibfnamefont {P.}~\bibnamefont {Combe}}, \bibinfo {editor}
		{\bibfnamefont {R.}~\bibnamefont {H{\o}egh-Krohn}}, \bibinfo {editor}
		{\bibfnamefont {G.}~\bibnamefont {Rideau}}, \bibinfo {editor} {\bibfnamefont
			{M.}~\bibnamefont {Sirugue-Collin}}, \bibinfo {editor} {\bibfnamefont
			{M.}~\bibnamefont {Sirugue}},\ and\ \bibinfo {editor} {\bibfnamefont
			{R.}~\bibnamefont {Stora}}}\ (\bibinfo  {publisher} {Springer Berlin
		Heidelberg},\ \bibinfo {address} {Berlin, Heidelberg},\ \bibinfo {year}
	{1979})\ pp.\ \bibinfo {pages} {337--360}\BibitemShut {NoStop}%
	\bibitem [{\citenamefont {Voros}(1983)}]{V83}%
	\BibitemOpen
	\bibfield  {author} {\bibinfo {author} {\bibfnamefont {A.}~\bibnamefont
			{Voros}},\ }\bibfield  {title} {\bibinfo {title} {The return of the quartic
			oscillator. the complex wkb method},\ }\href
	{http://www.numdam.org/item/AIHPA_1983__39_3_211_0} {\bibfield  {journal}
		{\bibinfo  {journal} {Annales de l'I.H.P. Physique th\'eorique}\ }\textbf
		{\bibinfo {volume} {39}},\ \bibinfo {pages} {211} (\bibinfo {year}
		{1983})}\BibitemShut {NoStop}%
	\bibitem [{\citenamefont {Balian}\ \emph {et~al.}(1978)\citenamefont {Balian},
		\citenamefont {Parisi},\ and\ \citenamefont {Voros}}]{BPV78}%
	\BibitemOpen
	\bibfield  {author} {\bibinfo {author} {\bibfnamefont {R.}~\bibnamefont
			{Balian}}, \bibinfo {author} {\bibfnamefont {G.}~\bibnamefont {Parisi}},\
		and\ \bibinfo {author} {\bibfnamefont {A.}~\bibnamefont {Voros}},\ }\bibfield
	{title} {\bibinfo {title} {Discrepancies from asymptotic series and their
			relation to complex classical trajectories},\ }\href
	{https://doi.org/10.1103/PhysRevLett.41.1141} {\bibfield  {journal} {\bibinfo
			{journal} {Phys. Rev. Lett.}\ }\textbf {\bibinfo {volume} {41}},\ \bibinfo
		{pages} {1141} (\bibinfo {year} {1978})}\BibitemShut {NoStop}%
	\bibitem [{\citenamefont {Okun}\ and\ \citenamefont
		{Burke}(2021{\natexlab{b}})}]{OB21b}%
	\BibitemOpen
	\bibfield  {author} {\bibinfo {author} {\bibfnamefont {P.}~\bibnamefont
			{Okun}}\ and\ \bibinfo {author} {\bibfnamefont {K.}~\bibnamefont {Burke}},\
	}\bibfield  {title} {\bibinfo {title} {Uncommonly accurate energies for the
			general quartic oscillator},\ }\href
	{https://doi.org/https://doi.org/10.1002/qua.26554} {\bibfield  {journal}
		{\bibinfo  {journal} {International Journal of Quantum Chemistry}\ }\textbf
		{\bibinfo {volume} {121}},\ \bibinfo {pages} {e26554} (\bibinfo {year}
		{2021}{\natexlab{b}})}\BibitemShut {NoStop}%
	\bibitem [{\citenamefont {Blinder}(2019)}]{B19}%
	\BibitemOpen
	\bibfield  {author} {\bibinfo {author} {\bibfnamefont {S.~M.}\ \bibnamefont
			{Blinder}},\ }\bibfield  {title} {\bibinfo {title} {Eigenvalues for a pure
			quartic oscillator}} (\bibinfo {year} {2019}),\ \bibinfo {note}
	{arXiv:1903.07471}\BibitemShut {NoStop}%
	\bibitem [{\citenamefont {Reid}(1970)}]{R70}%
	\BibitemOpen
	\bibfield  {author} {\bibinfo {author} {\bibfnamefont {C.~E.}\ \bibnamefont
			{Reid}},\ }\bibfield  {title} {\bibinfo {title} {Energy eigenvalues and
			matrix elements for the quartic oscillator},\ }\href
	{https://doi.org/https://doi.org/10.1016/0022-2852(70)90103-7} {\bibfield
		{journal} {\bibinfo  {journal} {Journal of Molecular Spectroscopy}\ }\textbf
		{\bibinfo {volume} {36}},\ \bibinfo {pages} {183 } (\bibinfo {year}
		{1970})}\BibitemShut {NoStop}%
	\bibitem [{\citenamefont {Bender}\ \emph {et~al.}(1977)\citenamefont {Bender},
		\citenamefont {Olaussen},\ and\ \citenamefont {Wang}}]{BOW77}%
	\BibitemOpen
	\bibfield  {author} {\bibinfo {author} {\bibfnamefont {C.~M.}\ \bibnamefont
			{Bender}}, \bibinfo {author} {\bibfnamefont {K.}~\bibnamefont {Olaussen}},\
		and\ \bibinfo {author} {\bibfnamefont {P.~S.}\ \bibnamefont {Wang}},\
	}\bibfield  {title} {\bibinfo {title} {Numerological analysis of the wkb
			approximation in large order},\ }\href
	{https://doi.org/10.1103/PhysRevD.16.1740} {\bibfield  {journal} {\bibinfo
			{journal} {Phys. Rev. D}\ }\textbf {\bibinfo {volume} {16}},\ \bibinfo
		{pages} {1740} (\bibinfo {year} {1977})}\BibitemShut {NoStop}%
	\bibitem [{\citenamefont {Voros}(1980)}]{V80}%
	\BibitemOpen
	\bibfield  {author} {\bibinfo {author} {\bibfnamefont {A.}~\bibnamefont
			{Voros}},\ }\bibfield  {title} {\bibinfo {title} {The zeta function of the
			quartic oscillator},\ }\href
	{https://doi.org/https://doi.org/10.1016/0550-3213(80)90085-1} {\bibfield
		{journal} {\bibinfo  {journal} {Nuclear Physics B}\ }\textbf {\bibinfo
			{volume} {165}},\ \bibinfo {pages} {209 } (\bibinfo {year}
		{1980})}\BibitemShut {NoStop}%
	\bibitem [{\citenamefont {Delabaere}\ and\ \citenamefont {Pham}(1997)}]{DP97}%
	\BibitemOpen
	\bibfield  {author} {\bibinfo {author} {\bibfnamefont {E.}~\bibnamefont
			{Delabaere}}\ and\ \bibinfo {author} {\bibfnamefont {F.}~\bibnamefont
			{Pham}},\ }\bibfield  {title} {\bibinfo {title} {Unfolding the quartic
			oscillator},\ }\href {https://doi.org/https://doi.org/10.1006/aphy.1997.5737}
	{\bibfield  {journal} {\bibinfo  {journal} {Annals of Physics}\ }\textbf
		{\bibinfo {volume} {261}},\ \bibinfo {pages} {180 } (\bibinfo {year}
		{1997})}\BibitemShut {NoStop}%
	\bibitem [{\citenamefont {White}\ and\ \citenamefont {Kutlin}(2017)}]{WK17}%
	\BibitemOpen
	\bibfield  {author} {\bibinfo {author} {\bibfnamefont {R.~B.}\ \bibnamefont
			{White}}\ and\ \bibinfo {author} {\bibfnamefont {A.~G.}\ \bibnamefont
			{Kutlin}},\ }\bibfield  {title} {\bibinfo {title} {Bound state energies using
			phase integral analysis}} (\bibinfo {year} {2017}),\ \bibinfo {note}
	{arXiv:1704.01170}\BibitemShut {NoStop}%
	\bibitem [{\citenamefont {Dunne}\ and\ \citenamefont {\"Unsal}(2014)}]{DU14}%
	\BibitemOpen
	\bibfield  {author} {\bibinfo {author} {\bibfnamefont {G.~V.}\ \bibnamefont
			{Dunne}}\ and\ \bibinfo {author} {\bibfnamefont {M.}~\bibnamefont
			{\"Unsal}},\ }\bibfield  {title} {\bibinfo {title} {Uniform wkb,
			multi-instantons, and resurgent trans-series},\ }\href
	{https://doi.org/10.1103/PhysRevD.89.105009} {\bibfield  {journal} {\bibinfo
			{journal} {Phys. Rev. D}\ }\textbf {\bibinfo {volume} {89}},\ \bibinfo
		{pages} {105009} (\bibinfo {year} {2014})}\BibitemShut {NoStop}%
	\bibitem [{\citenamefont {Dunne}\ and\ \citenamefont {{\"U}nsal}(2017)}]{DU17}%
	\BibitemOpen
	\bibfield  {author} {\bibinfo {author} {\bibfnamefont {G.~V.}\ \bibnamefont
			{Dunne}}\ and\ \bibinfo {author} {\bibfnamefont {M.}~\bibnamefont
			{{\"U}nsal}},\ }\bibfield  {title} {\bibinfo {title} {Wkb and resurgence in
			the mathieu equation},\ }in\ \href@noop {} {\emph {\bibinfo {booktitle}
			{Resurgence, Physics and Numbers}}},\ \bibinfo {editor} {edited by\ \bibinfo
		{editor} {\bibfnamefont {F.}~\bibnamefont {Fauvet}}, \bibinfo {editor}
		{\bibfnamefont {D.}~\bibnamefont {Manchon}}, \bibinfo {editor} {\bibfnamefont
			{S.}~\bibnamefont {Marmi}},\ and\ \bibinfo {editor} {\bibfnamefont
			{D.}~\bibnamefont {Sauzin}}}\ (\bibinfo  {publisher} {Scuola Normale
		Superiore},\ \bibinfo {address} {Pisa},\ \bibinfo {year} {2017})\ pp.\
	\bibinfo {pages} {249--298}\BibitemShut {NoStop}%
	\bibitem [{\citenamefont {Delabaere}\ \emph {et~al.}(1997)\citenamefont
		{Delabaere}, \citenamefont {Dillinger},\ and\ \citenamefont {Pham}}]{DDP97}%
	\BibitemOpen
	\bibfield  {author} {\bibinfo {author} {\bibfnamefont {E.}~\bibnamefont
			{Delabaere}}, \bibinfo {author} {\bibfnamefont {H.}~\bibnamefont
			{Dillinger}},\ and\ \bibinfo {author} {\bibfnamefont {F.}~\bibnamefont
			{Pham}},\ }\bibfield  {title} {\bibinfo {title} {Exact semiclassical
			expansions for one-dimensional quantum oscillators},\ }\href
	{https://doi.org/10.1063/1.532206} {\bibfield  {journal} {\bibinfo  {journal}
			{Journal of Mathematical Physics}\ }\textbf {\bibinfo {volume} {38}},\
		\bibinfo {pages} {6126} (\bibinfo {year} {1997})}\BibitemShut {NoStop}%
	\bibitem [{\citenamefont {{Hua}}(2012)}]{H12}%
	\BibitemOpen
	\bibfield  {author} {\bibinfo {author} {\bibfnamefont {L.-K.}\ \bibnamefont
			{{Hua}}},\ }\href@noop {} {\emph {\bibinfo {title} {Introduction to Higher
				Mathematics}}}\ (\bibinfo  {publisher} {Cambridge University Press},\
	\bibinfo {address} {Cambridge, UK},\ \bibinfo {year} {2012})\BibitemShut
	{NoStop}%
	\bibitem [{\citenamefont {Kohn}\ and\ \citenamefont {Sham}(1965)}]{KS65}%
	\BibitemOpen
	\bibfield  {author} {\bibinfo {author} {\bibfnamefont {W.}~\bibnamefont
			{Kohn}}\ and\ \bibinfo {author} {\bibfnamefont {L.~J.}\ \bibnamefont
			{Sham}},\ }\bibfield  {title} {\bibinfo {title} {Self-consistent equations
			including exchange and correlation effects},\ }\href
	{https://doi.org/10.1103/PhysRev.140.A1133} {\bibfield  {journal} {\bibinfo
			{journal} {Phys. Rev.}\ }\textbf {\bibinfo {volume} {140}},\ \bibinfo {pages}
		{A1133} (\bibinfo {year} {1965})}\BibitemShut {NoStop}%
	\bibitem [{\citenamefont {Burke}\ and\ \citenamefont {Wagner}(2013)}]{WB13}%
	\BibitemOpen
	\bibfield  {author} {\bibinfo {author} {\bibfnamefont {K.}~\bibnamefont
			{Burke}}\ and\ \bibinfo {author} {\bibfnamefont {L.~O.}\ \bibnamefont
			{Wagner}},\ }\bibfield  {title} {\bibinfo {title} {Dft in a nutshell},\
	}\href {https://doi.org/https://doi.org/10.1002/qua.24259} {\bibfield
		{journal} {\bibinfo  {journal} {International Journal of Quantum Chemistry}\
		}\textbf {\bibinfo {volume} {113}},\ \bibinfo {pages} {96} (\bibinfo {year}
		{2013})}\BibitemShut {NoStop}%
	\bibitem [{\citenamefont {Ribeiro}\ \emph {et~al.}(2015)\citenamefont
		{Ribeiro}, \citenamefont {Lee}, \citenamefont {Cangi}, \citenamefont
		{Elliott},\ and\ \citenamefont {Burke}}]{RLCE15}%
	\BibitemOpen
	\bibfield  {author} {\bibinfo {author} {\bibfnamefont {R.~F.}\ \bibnamefont
			{Ribeiro}}, \bibinfo {author} {\bibfnamefont {D.}~\bibnamefont {Lee}},
		\bibinfo {author} {\bibfnamefont {A.}~\bibnamefont {Cangi}}, \bibinfo
		{author} {\bibfnamefont {P.}~\bibnamefont {Elliott}},\ and\ \bibinfo {author}
		{\bibfnamefont {K.}~\bibnamefont {Burke}},\ }\bibfield  {title} {\bibinfo
		{title} {Corrections to thomas-fermi densities at turning points and
			beyond},\ }\href {https://doi.org/10.1103/PhysRevLett.114.050401} {\bibfield
		{journal} {\bibinfo  {journal} {Phys. Rev. Lett.}\ }\textbf {\bibinfo
			{volume} {114}},\ \bibinfo {pages} {050401} (\bibinfo {year}
		{2015})}\BibitemShut {NoStop}%
	\bibitem [{\citenamefont {Ribeiro}\ and\ \citenamefont {Burke}(2017)}]{RB17}%
	\BibitemOpen
	\bibfield  {author} {\bibinfo {author} {\bibfnamefont {R.~F.}\ \bibnamefont
			{Ribeiro}}\ and\ \bibinfo {author} {\bibfnamefont {K.}~\bibnamefont
			{Burke}},\ }\bibfield  {title} {\bibinfo {title} {Leading corrections to
			local approximations. ii. the case with turning points},\ }\href
	{https://doi.org/10.1103/PhysRevB.95.115115} {\bibfield  {journal} {\bibinfo
			{journal} {Phys. Rev. B}\ }\textbf {\bibinfo {volume} {95}},\ \bibinfo
		{pages} {115115} (\bibinfo {year} {2017})}\BibitemShut {NoStop}%
	\bibitem [{\citenamefont {Ribeiro}\ and\ \citenamefont {Burke}(2018)}]{RB18}%
	\BibitemOpen
	\bibfield  {author} {\bibinfo {author} {\bibfnamefont {R.~F.}\ \bibnamefont
			{Ribeiro}}\ and\ \bibinfo {author} {\bibfnamefont {K.}~\bibnamefont
			{Burke}},\ }\bibfield  {title} {\bibinfo {title} {Deriving uniform
			semiclassical approximations for one-dimensional fermionic systems},\ }\href
	{https://doi.org/10.1063/1.5025628} {\bibfield  {journal} {\bibinfo
			{journal} {J. Chem. Phys.}\ }\textbf {\bibinfo {volume} {148}},\ \bibinfo
		{pages} {194103} (\bibinfo {year} {2018})}\BibitemShut {NoStop}%
	\bibitem [{\citenamefont {Burke}(2007)}]{ABC}%
	\BibitemOpen
	\bibfield  {author} {\bibinfo {author} {\bibfnamefont {K.}~\bibnamefont
			{Burke}},\ }\bibfield  {title} {\bibinfo {title} {The abc of dft}} (\bibinfo
	{year} {2007}),\ \bibinfo {note}
	{\url{https://dft.uci.edu/doc/g1.pdf}}\BibitemShut {NoStop}%
	\bibitem [{\citenamefont {White}(1992)}]{W92}%
	\BibitemOpen
	\bibfield  {author} {\bibinfo {author} {\bibfnamefont {S.~R.}\ \bibnamefont
			{White}},\ }\bibfield  {title} {\bibinfo {title} {Density matrix formulation
			for quantum renormalization groups},\ }\href
	{https://doi.org/10.1103/PhysRevLett.69.2863} {\bibfield  {journal} {\bibinfo
			{journal} {Phys. Rev. Lett.}\ }\textbf {\bibinfo {volume} {69}},\ \bibinfo
		{pages} {2863} (\bibinfo {year} {1992})}\BibitemShut {NoStop}%
	\bibitem [{\citenamefont {Pilati}\ \emph {et~al.}(2018)\citenamefont {Pilati},
		\citenamefont {Zintchenko}, \citenamefont {Troyer},\ and\ \citenamefont
		{Ancilotto}}]{PZTA18}%
	\BibitemOpen
	\bibfield  {author} {\bibinfo {author} {\bibfnamefont {S.}~\bibnamefont
			{Pilati}}, \bibinfo {author} {\bibfnamefont {I.}~\bibnamefont {Zintchenko}},
		\bibinfo {author} {\bibfnamefont {M.}~\bibnamefont {Troyer}},\ and\ \bibinfo
		{author} {\bibfnamefont {F.}~\bibnamefont {Ancilotto}},\ }\bibfield  {title}
	{\bibinfo {title} {Density functional theory versus quantum monte carlo
			simulations of fermi gases in the optical-lattice arena},\ }\href
	{https://doi.org/10.1140/epjb/e2018-90021-1} {\bibfield  {journal} {\bibinfo
			{journal} {The European Physical Journal B}\ ,\ \bibinfo {pages} {70}}
		(\bibinfo {year} {2018})}\BibitemShut {NoStop}%
	\bibitem [{\citenamefont {Stoudenmire}\ \emph {et~al.}(2012)\citenamefont
		{Stoudenmire}, \citenamefont {Wagner}, \citenamefont {White},\ and\
		\citenamefont {Burke}}]{SWWB12}%
	\BibitemOpen
	\bibfield  {author} {\bibinfo {author} {\bibfnamefont {E.~M.}\ \bibnamefont
			{Stoudenmire}}, \bibinfo {author} {\bibfnamefont {L.~O.}\ \bibnamefont
			{Wagner}}, \bibinfo {author} {\bibfnamefont {S.~R.}\ \bibnamefont {White}},\
		and\ \bibinfo {author} {\bibfnamefont {K.}~\bibnamefont {Burke}},\ }\bibfield
	{title} {\bibinfo {title} {One-dimensional continuum electronic structure
			with the density-matrix renormalization group and its implications for
			density-functional theory},\ }\href
	{https://doi.org/10.1103/PhysRevLett.109.056402} {\bibfield  {journal}
		{\bibinfo  {journal} {Phys. Rev. Lett.}\ }\textbf {\bibinfo {volume} {109}},\
		\bibinfo {pages} {056402} (\bibinfo {year} {2012})}\BibitemShut {NoStop}%
	\bibitem [{\citenamefont {Baker}\ \emph {et~al.}(2015)\citenamefont {Baker},
		\citenamefont {Stoudenmire}, \citenamefont {Wagner}, \citenamefont {Burke},\
		and\ \citenamefont {White}}]{BSWB15}%
	\BibitemOpen
	\bibfield  {author} {\bibinfo {author} {\bibfnamefont {T.~E.}\ \bibnamefont
			{Baker}}, \bibinfo {author} {\bibfnamefont {E.~M.}\ \bibnamefont
			{Stoudenmire}}, \bibinfo {author} {\bibfnamefont {L.~O.}\ \bibnamefont
			{Wagner}}, \bibinfo {author} {\bibfnamefont {K.}~\bibnamefont {Burke}},\ and\
		\bibinfo {author} {\bibfnamefont {S.~R.}\ \bibnamefont {White}},\ }\bibfield
	{title} {\bibinfo {title} {One-dimensional mimicking of electronic structure:
			The case for exponentials},\ }\href
	{https://doi.org/10.1103/PhysRevB.91.235141} {\bibfield  {journal} {\bibinfo
			{journal} {Phys. Rev. B}\ }\textbf {\bibinfo {volume} {91}},\ \bibinfo
		{pages} {235141} (\bibinfo {year} {2015})}\BibitemShut {NoStop}%
	\bibitem [{\citenamefont {Elliott}\ \emph {et~al.}(2015)\citenamefont
		{Elliott}, \citenamefont {Cangi}, \citenamefont {Pittalis}, \citenamefont
		{Gross},\ and\ \citenamefont {Burke}}]{ECPG15}%
	\BibitemOpen
	\bibfield  {author} {\bibinfo {author} {\bibfnamefont {P.}~\bibnamefont
			{Elliott}}, \bibinfo {author} {\bibfnamefont {A.}~\bibnamefont {Cangi}},
		\bibinfo {author} {\bibfnamefont {S.}~\bibnamefont {Pittalis}}, \bibinfo
		{author} {\bibfnamefont {E.~K.~U.}\ \bibnamefont {Gross}},\ and\ \bibinfo
		{author} {\bibfnamefont {K.}~\bibnamefont {Burke}},\ }\bibfield  {title}
	{\bibinfo {title} {Almost exact exchange at almost no computational cost in
			electronic structure},\ }\href {https://doi.org/10.1103/PhysRevA.92.022513}
	{\bibfield  {journal} {\bibinfo  {journal} {Phys. Rev. A}\ }\textbf {\bibinfo
			{volume} {92}},\ \bibinfo {pages} {022513} (\bibinfo {year}
		{2015})}\BibitemShut {NoStop}%
\end{thebibliography}
